\documentclass{aa}  

\usepackage{graphicx}
\usepackage{xcolor}
\usepackage{txfonts}
\usepackage{multirow}

\usepackage[hidelinks,colorlinks=true,linkcolor=red,citecolor=blue,urlcolor=violet]{hyperref}
\usepackage{natbib}

\newcommand{\rhon}{\rho_{\rm n}}
\newcommand{\rs}{r_{\rm s}}

\newcommand{\rhogas}{\rho_{\rm gas}}
\newcommand{\rhogasstar}{\rho_{\rm gas,\star}}
\newcommand{\pstar}{p_\star}
\newcommand{\Tstar}{T_{\star}}
\newcommand{\nstar}{n_\star}
\newcommand{\neestar}{n_{\rm{e,}\star}}
\newcommand{\Rstar}{R_\star}
\newcommand{\zstar}{z_\star}
\newcommand{\kboltz}{k_{\rm B}}
\newcommand{\proton}{m_{\rm p}}
\newcommand{\gammain}{\gamma'_{\rm{IN}}}
\newcommand{\gammaout}{\gamma'_{\rm{OUT}}}
\newcommand{\Phieff}{\Phi_{\rm{eff}}}
\newcommand{\Mdue}{M_{\rm{200}}}
\newcommand{\rdue}{r_{\rm{200}}}
\newcommand{\Mcinque}{M_{\rm{500}}}
\newcommand{\rcinque}{r_{\rm{500}}}
\newcommand{\cdue}{c_{\rm{200}}}
\newcommand{\rhocrit}{\rho_{\rm{crit}}}
\newcommand{\rminus}{r_{\rm{-2}}}
\newcommand{\rhoshell}{\rho_{\rm{shell}}}
\newcommand{\Rbreak}{R_{\rm{break}}}
\newcommand{\uzero}{u_{\rm{0}}}
\newcommand{\Rzero}{R_{\rm{0}}}

\newcommand{\nee}{n_{\rm{e}}}

\newcommand{\Tsp}{T_{\rm{sp}}}

\newcommand{\nii}{n_{\rm{i}}}
\newcommand{\Tsl}{T_{\rm{sl}}}
\newcommand{\nshell}{n_{\rm{e,shell}}}
\newcommand{\Tmean}{T_{\rm{sl,mean}}}

\newcommand{\zeff}{z_{\rm{eff}}}
\newcommand{\sigmaTH}{\sigma_{\rm{TH}}}

\newcommand{\vlos}{v_{\rm{los}}}

\newcommand{\sigmaturb}{\sigma_{\rm{turb}}}
\newcommand{\Qin}{Q^{\rm{in}}}
\newcommand{\Qout}{Q^{\rm{out}}}
\newcommand{\errQ}{err_{\rm{Q}}}
\newcommand{\SsQ}{S_{\rm{Q}}}
\newcommand{\MHE}{M_{\rm{HE}}}
\newcommand{\Mrot}{M_{\rm{rot}}}
\newcommand{\Mtrue}{M_{\rm{true}}}

\newcommand{\Tgas}{T}
\newcommand{\rDelta}{r_{\rm{\Delta}}}
\newcommand{\rhogasshell}{\rho_{\rm{gas,shell}}}
\newcommand{\pshell}{p_{\rm{shell}}}
\newcommand{\MDelta}{M_{\rm{\Delta}}}
\newcommand{\cDelta}{c_{\rm{\Delta}}}

\begin{document} 
\defcitealias{lovisari20}{L20}
\defcitealias{ZHANG10}{Z10}
\defcitealias{MAHDAVI2013}{M13}

   \title{Gas rotation and dark matter halo shape in cool-core clusters of galaxies}

   \author{T. Bartalesi\inst{1}\fnmsep\thanks{\email{tommaso.bartalesi2@studio.unibo.it}}
          \and
          S. Ettori\inst{2,3}\fnmsep\thanks{\email{stefano.ettori@inaf.it}}
          \and
          C. Nipoti\inst{1}
          }

   \institute{Dipartimento di Fisica e Astronomia “Augusto Righi” – Alma Mater Studiorum – Università di Bologna, via Gobetti 93/2, I-40129 Bologna 
         \and INAF, Osservatorio di Astrofisica e Scienza dello Spazio, via Piero Gobetti 93/3, 40129 Bologna, Italy 
         \and INFN, Sezione di Bologna, viale Berti Pichat 6/2, 40127 Bologna, Italy
             }

\date{Accepted, October 24, 2023}
 
  \abstract
   {}
   {We study the possibility that the gas in cool-core clusters of galaxies has non-negligible rotation support, the impact of gas rotation on mass estimates from current X-ray observations, and the ability of forthcoming X-ray observatories to detect such rotation.}
   {We present three representative models of massive cool-core clusters with rotating intracluster medium (ICM) in equilibrium in cosmologically motivated spherical, oblate or prolate dark matter halos, represented by physical density-potential pairs.
   In the models, the gas follows a composite-polytropic distribution, and has rotation velocity profiles consistent with current observational constraints and similar to those found in clusters formed in cosmological simulations.
   We show that the models are consistent with the available measurements of the ICM properties of the massive cluster population: thermodynamic profiles, shape of surface-brightness distribution, hydrostatic mass bias and broadening of X-ray emitting lines. 
   Using the configuration for the microcalorimeter onboard the \emph{XRISM} satellite, we generate a set of mock X-ray spectra of our cluster models, which we then analyze to make predictions on the estimates of the rotation speed that will be obtained with such an instrument.
   We then assess what fraction of the hydrostatic mass bias of our models could be accounted for by detecting rotation speed with \emph{XRISM} spectroscopy over the range $(0.1-1)r_{500}$, sampled with 3 nonoverlapping pointings. 
   }
   {Current data leave room for rotating ICM in cool-core clusters with peaks of rotation speed as high as $600\,\mathrm{km/s}$. 
   We have shown that such rotation, if present, will be detected with upcoming X-ray facilities such as \emph{XRISM} and that $60-70\%$ of the hydrostatic mass bias due to rotation can be accounted for using the line-of-sight velocity measured from X-ray spectroscopy with \emph{XRISM}, with a residual bias smaller than $3\%$ at an overdensity of 500. In this way, \emph{XRISM} will allow us to pin down any mass bias of origin different from rotation.}
{}

   \maketitle
%

\section{Introduction}
\label{sec.Intro}
The mass of clusters of galaxies is crucial to understand the formation and evolution of cosmic structures, and to constrain the parameters that define the cosmological background (see \citealt{P19} for a review). 
Clusters of galaxies are permeated by a hot ($\sim 10^7-10^8$ K), rarefied  ($\sim 10^{-2}-10^{-4}$ particles per cm$^3$), optically thin, gaseous component, known as intracluster medium (ICM), which emits in the X-rays via thermal Bremsstrahlung and emission lines from collisional excitation of inner shell electrons of heavy metals. 
Assuming that the ICM is in hydrostatic equilibrium, X-ray observations can thus be used to infer the mass of galaxy clusters (see \citealt{E13} for a review). Mass estimates obtained in this way can be very precise, but not accurate \citep[e.g.][]{E19}, given that the hydrostatic equilibrium does not account for the residual non-thermalized (kinetic) energy in the ICM \citep[see e.g.][]{rasia06,piffaretti08,lau09,suto13,lau13,biffi16,angelinelli20,GIANFAGNA21}.
This effect that brings hydrostatic masses to underestimate the ''true'' mass is often referred to as hydrostatic mass bias.
Measurements of this bias can be obtained by comparison with more direct mass estimators \citep[e.g.][]{ZHANG10,MAHDAVI2013,lovisari20}. 
In particular, being the most massive gravitationally bound structures in the Universe, galaxy clusters are effective gravitational lenses, which provide a complementary, and typically more accurate method to infer the total (i.e. baryon plus dark matter) mass \citep[see e.g.][]{meneghetti10,rasia12}.
Alternatively, the dynamical mass of a cluster can be estimated by exploiting measurements of the orbital velocities of its member galaxies  \citep[see e.g.][]{ferragamo21}. 

Even though in the X-ray observations the gas clumpiness and the use of the spectroscopic temperature in reconstructing the thermal properties of the ICM contributed nonnegligibly to the hydrostatic mass bias \citep[see e.g.][]{rasia06,roncarelli13,pearce20,Towler23}, most of this bias is expected to be due to the motions in the ICM: in particular, turbulence bulk motion, and rotation \citep[see e.g.][]{N07,nelson14,biffi16,angelinelli20}.
Most of these previous works have focused on the relative importance of bulk and random motions to the total budget of the hydrostatic mass bias, with only a few studies dedicated to the contribution from the ICM rotational support \citep[e.g.][]{F09}. 
There are essentially only two direct ways to measure gas rotation in galaxy clusters: the rotational kinetic Sunyaev-Zeldovich effect (\citealt{C02}, \citealt{Chluba02} and also \citealt{Sunyaev80}; see \citealt{B18} and \citealt{A23} for future perspectives)
and the Doppler shift of the centroids of the X-ray emitting lines or their Doppler broadening. 
The latter measurements require X-ray spectrometers at high energy resolution ($\Delta E \lesssim 10\,\mathrm{eV}$ at $E \approx 6-7\,\mathrm{keV}$ is required to detect a line-of-sight speed of $\approx 500\,\mathrm{km/s}$; e.g. \citealt{S03}, \citealt{B13}), which are thus far reached only by calorimeter onboard International X-ray Astronomy Mission \emph{ASTRO-H/Hitomi}\footnote{See \href{https://www.isas.jaxa.jp/en/missions/spacecraft/past/hitomi.html}{https://www.isas.jaxa.jp/en/missions/spacecraft/past/hitomi.html}.} satellite (see \citealt{H16} for its results). 
The loss of \emph{Hitomi} have prevented us from depicting a comprehensive overview of the kinematics of the ICM; however, the forthcoming microcalorimeter \emph{Resolve} onboard the X-Ray Imaging and Spectroscopy Mission\footnote{See \href{https://xrism.isas.jaxa.jp/en/}{https://xrism.isas.jaxa.jp/en/}.} (\emph{XRISM}) satellite (with $\Delta E\simeq 7\,\mathrm{eV}$ FWHM at $E=6-7\,\mathrm{keV}$), launched in September 2023, is expected to provide some key elements to improve our understanding of the ICM kinematics. 
Nowadays, only upper limits on the velocity broadening of X-ray emitting lines are available: using X-ray Multi-Mirror Mission\footnote{See \href{https://www.cosmos.esa.int/web/xmm-newton}{https://www.cosmos.esa.int/web/xmm-newton}.}  (\emph{XMM-Newton}) Reflection Grating Spectrometers (RGS) data, \cite{P15}, in most cool cores of clusters, groups, and massive elliptical galaxies of their observed sample, found broadening velocities of $\approx 500\,\mathrm{km/s}$ (see also \citealt{S11} and \citealt{BAMBIC}). Even though some objects have higher upper limits (of $\approx 1000\,\mathrm{km/s}$), we interpret $500\,\mathrm{km/s}$ as the current upper limit on the rotation speed of the ICM in typical clusters, which leaves open the possibility that the ICM has nonnegligible rotation support in relaxed clusters\footnote{Indications of rotation support of the galactic component have been found in some clusters from spectroscopic observations of member galaxies \citep[see e.g.][]{Oegerle92, Hwang07, Ferrami23}. 
The differences in the rotation speed profiles of the ICM and member galaxies are an interesting issue to be explored with future facilities.}.

In the cosmological context, the rotation of both dark matter (DM) and gas is expected to be induced primarily by the large-scale processes involving the entire cluster (such as tidal torques from neighbouring overdensities; \citealt{P69}). In massive clusters (virial masses $\gtrsim 5\times 10^{14}\,\mathrm{M_\odot}$) formed in cosmological $N$-body hydrodynamical nonradiative simulations, \cite{BALDI17}  have found that the rotation support of the ICM tends to be higher than that of the DM, with values of the gas spin parameter on average  higher by $13\%$ than those of the halo spin parameter. 
In principle, the rotation support of the ICM can be further enhanced by unimpeded radiative cooling, because of conservation of angular momentum \cite[see e.g.][]{Kley95}, but in real clusters also heating mechanisms are at work.
In fact, including radiative cooling, Active Galactic Nucleus (AGN) and stellar feedback models in cosmological simulations, \cite{BALDI17} have found that the rotation support of the ICM is similar to that found in nonradiative simulations. 

Based on the properties of the ICM in the central regions, clusters of galaxies are classified as cool-core and non cool-core clusters \citep[e.g.\ section 6.4.3 of][]{C19}.
Given that we are interested in rotation support of the ICM, in this work we focus on cool-core clusters, which tend to be relaxed \citep[e.g.][]{Pratt10,MAHDAVI2013} and thus good targets for symmetric equilibrium models of the ICM. 
By definition, cool-core clusters are characterized by lower central ICM entropy,
which is broadly interpreted as a signature of cooling. In fact, the measured values of the central entropy are much higher than predicted in a standard cooling-flow model (e.g.\ \citealt{M18}).
This suggests that, in time-averaged sense, over $\sim 10\,\mathrm{Gyr}$, radiative cooling is balanced by some form of heating, a picture also supported by the fact that radiative cosmological simulations without heating suffer from the "overcooling" problem, which produces photometric features inconsistent with observations \citep[e.g.][]{F09,L11,Lau12,NA13}.
There is growing consensus that AGN feedback provides the dominant heating contribution in the cluster inner regions (see \citealt{MN12, L22} for reviews  and \citealt{Nobels22,Husko22} for recent results).
However, it must be stressed that modeling the complex interplay of heating and cooling is challenging also for state-of-the art simulations. For instance, clusters formed in currently available cosmological simulations including an AGN feedback model can suffer from the "entropy-core" problem, in the sense that their inner entropy profiles do not match those observed in real clusters \citep{Altamura23AGN}.

Rotation of the ICM could be relevant also to the energy balance of cool cores, given that the ICM is known to be weakly magnetized.   
If the magnetized, rotating ICM is unstable to  the magnetorotational instability \citep{Balbus1991}, the nonlinear evolution of the instability will lead to turbulent heating, which could contribute to offsetting the radiative cooling of the ICM and to halting the cooling flows, lending a hand to the AGN feedback
(see \citealt{Nipoti2014,N15}).

In this work, we propose three models representative of typical, nearby, massive cool-core clusters, with cosmologically motivated dark halos with different shapes (Sect. \ref{sec.2}) and rotating ICM with rotation speed consistent with observational upper limits (Sect. \ref{sec.3}). 
In Sect. \ref{sec:comparison}, we compare the intrinsic and observable properties of the ICM in our cluster models to the observational data of real galaxy clusters. 
In Sect. \ref{sec.DETECT}, we assess the detectability of the rotation support of our models, building mock X-ray spectra of the rotating ICM in our cluster models, using the configurations for \emph{Resolve}. Sect. \ref{sec.CONCL} concludes. 

Throughout this article, when using the Hubble parameter $H(z)=H_0E^{1/2}(z)$, where $E(z)=\sqrt{\Omega_{\Lambda,0}+\Omega_{\mathrm{m},0}(1+z)^3}$, we assume $\Omega_{\mathrm{m},0}=0.3$, $\Omega_{\Lambda,0}=0.7$ and Hubble constant $H_0=70\,\mathrm{km/s/Mpc}$.

\section{Dark matter halo models}
\label{sec.2}

We introduce here the gravitational potentials that we will use to build our cluster models. Given that the mass content of clusters is dominated by the dark matter (DM), these gravitational potentials must be essentially representative of those produced by the cluster DM halos. 

Cosmological $N$-body DM-only simulations predict for most halos an aspherical shape, set at the time of the last major merger (\citealt{A06}).
In general, the angle-averaged density profile of these simulated halos is well fitted by the Navarro-Frenk-White (NFW; \citealt{NFW96}) profile
\begin{equation}
\label{eq.NFW}
\rho(r)=\frac{\rhon}{\left(\frac{r}{\rs}\right)\left( 1+\frac{r}{\rs}\right)^2},
\end{equation}
where $r$ is the distance from the halo center, $\rhon$ is a characteristic density and $\rs$ is the scale radius.
The density distribution of DM in real clusters is also well represented by this profile: for instance, from X-ray and Sunyaev-Zeldovich effect observations, \cite{E19} infer that the NFW profile successfully models the angle averaged density profiles of the halos of the observed clusters.
It is thus natural to take the NFW density profile (Eq.\ \ref{eq.NFW}) as reference to build realistic flattened halo models.
In the following sections we will describe how we build axisymmetric halo models by suitably modifying the spherical NFW model.

 \begin{figure*}
   \centering
   \includegraphics[width=0.49\textwidth]{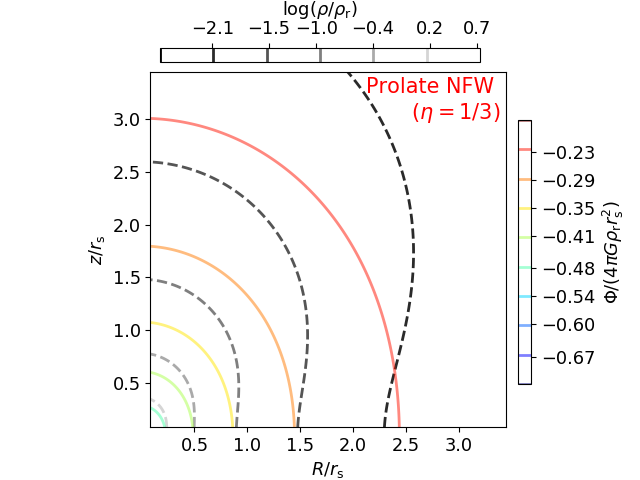}
   \includegraphics[width=0.49\textwidth]{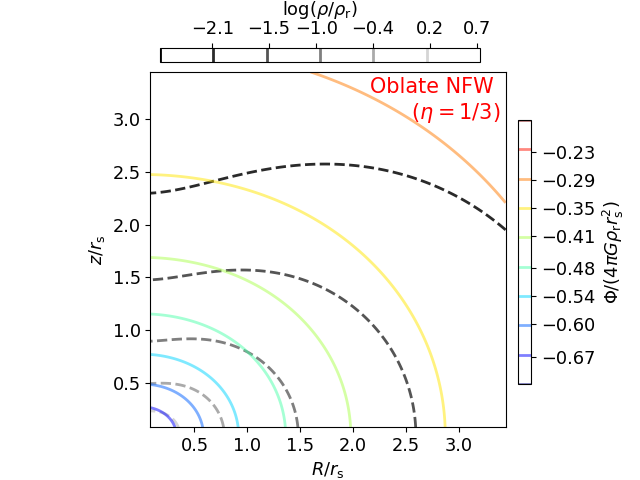}
   \caption{Isodensity (dashed) and isopotential (solid) contours in the meridional plane of the prolate (left panel) and oblate (right panel) NFW models with $\eta=1/3$. The size of the box is  $\approx\rdue/\rs$ (see Sect. \ref{sec.2.2}).}
   \label{fig.DMM}
    \end{figure*}

\subsection{Flattened NFW density-potential pairs}
\label{sec.2.1}

\citet{C05} presented a technique to construct analytic axisymmetric and triaxial density-potential pairs by modifying a parent spherical density distribution with given density profile $\tilde\rho(\tilde r)$,  where $\tilde\rho=\rho/\rhon$ and $\tilde{r}=r/\rs$, with $\rhon$ a characteristic density and $\rs$ a scale radius.
The generic density-potential pair of this family can be written in Cartesian coordinates $(x,y,z)$ as
\begin{equation}
\label{eq.DC}
\tilde{\rho}(x,y,z) = \tilde{\rho}(\tilde{r}) + \frac{\epsilon \tilde{y}^2 + \eta \tilde{z}^2}{\tilde{r}} \rho'(\tilde{r}),
\end{equation}
where $\tilde x=x/\rs$, $\tilde y=y/\rs$, $\tilde z=z/\rs$, $\tilde r=\sqrt{\tilde x^2+\tilde y^2+\tilde z^2}$ and $\rho' (\tilde{r})= {d\tilde \rho}/{d\tilde r}$, and 
\begin{equation}
\label{eq.PC}
    \tilde{\Phi}(x,y,z) = \tilde{\Phi}_0 (\tilde{r}) + (\epsilon + \eta) [\tilde{\Phi}_1 (\tilde{r}) - \tilde{\Phi}_0 (\tilde{r})] + (\epsilon \tilde{y}^2 + \eta \tilde{z}^2) \tilde{\Phi}_2 (\tilde{r}),
\end{equation}
where $\tilde\Phi=\Phi/(4\pi G\rhon\rs^2)$, $\Phi$ is the gravitational potential, and 
$\tilde{\Phi}_0$, $\tilde{\Phi}_1$ and $\tilde{\Phi}_2$ are functions depending on $\tilde\rho(r)$, whose definitions can be found in \citet{C05}. Here $\epsilon>0$ and $\eta>0$ are dimensionless parameters which must be such that $\tilde{\rho}(x,y,z)>0$ everywhere. We note that, though constructed exploiting the technique of the homeoidal expansion, the density-potential pairs given by the above formulae do not require $\epsilon$ and $\eta$ to be much smaller than unity \citep[see section 2 of][]{C05}.

Here we assume as parent spherical density profile the NFW model (Eq.\ \ref{eq.NFW}), which in dimensionless form reads 
\begin{equation}
\label{eq.NFW_tris}
\tilde \rho(\tilde r)=\frac{1}{\tilde r\left(1+\tilde r\right)^2}.
\end{equation}
Using Equation~(\ref{eq.NFW_tris}) as $\tilde\rho$, Eq. (\ref{eq.DC}) becomes
\begin{equation}
\label{eq.NFW1+}
\tilde{\rho}(x, y, z) = \frac{1}{\tilde{r}( 1+\tilde{r})^2} - \frac{\epsilon \tilde{y}^2  + \eta \tilde{z}^2}{\tilde{r}} \frac{1+3\tilde{r}}{\tilde{r}^2 (1+\tilde{r})^3}.
\end{equation}
The dimensionless gravitational potential generated by the density profile (\ref{eq.NFW1+}) is given by Eq. (\ref{eq.PC}), where
\begin{equation}
\label{eq.Phi0}
\tilde{\Phi}_0 (r) = - \frac{\ln(1+\tilde{r})}{\tilde{r}},
\end{equation}
\begin{equation}
\label{eq.Phi1}
\tilde{\Phi}_1 (\tilde{r}) = - \frac{1}{6 \tilde{r}} + \frac{2}{3 \tilde{r}^2} - \frac{\ln(1+\tilde{r})}{\tilde{r}^3} - \frac{1}{3 \tilde{r}^3 (\tilde{r} +1)} + \frac{1}{3 \tilde{r}^3} - \frac{1}{3 (1+\tilde{r})}
\end{equation}
and 
\begin{equation}
\label{eq.Phi2}
\tilde{\Phi}_2 (\tilde{r}) =  \frac{1}{2 \tilde{r}^3} - \frac{2}{ \tilde{r}^4} + \frac{3 \ln(1+\tilde{r})}{\tilde{r}^5} + \frac{1}{ \tilde{r}^5 (\tilde{r} +1)} - \frac{1}{\tilde{r}^5}.
\end{equation} 
The second term in the r.h.s.\ of Eq.\ (\ref{eq.NFW1+}) breaks the spherical symmetry of the distribution, subtracting density along the directions $\tilde y$ and $\tilde z$. It is evident that the dimensionless density distribution (\ref{eq.NFW1+}) would assume negative values if the directional subtraction of parent density is sufficiently large. When we consider the NFW as the parent density profile, the condition that at any point of space $\tilde{\rho}>0$, with $\tilde{\rho}$ given by Eq.\ (\ref{eq.NFW1+}),
imposes $\epsilon,\eta \leq1/3$ (see \citealt{C05} for the method to limit $\epsilon$ and $\eta$).

In particular, in this work we will consider prolate ($\eta=\epsilon$) and oblate ($\epsilon=0$) axisymmetric density-potential pairs, having as parent density distribution Eq.\ (\ref{eq.NFW_tris}), which we will refer to as prolate NFW and oblate NFW models, respectively.  
The prolate NFW model ($\eta=\epsilon$), renaming $x$ as $z$, and viceversa, has density distribution
 \begin{equation}
\label{eq.DP}
\tilde{\rho}(R,z) = \frac{1}{\tilde{r}( 1+\tilde{r})^2} - \frac{\eta \tilde{R}^2}{\tilde{r}} \frac{1+3\tilde{r}}{\tilde{r}^2 (1+\tilde{r})^3}\quad\mbox{(prolate)},
\end{equation}
(shown for $\eta=1/3$ in the left panel of Fig.\ \ref{fig.DMM}) and gravitational potential 
\begin{equation}\begin{split}
\label{eq.PP}
\tilde{\Phi}(R,z) =&  - \frac{\ln(1+\tilde{r})}{\tilde{r}} +2\eta \left[-\frac{1}{6 \tilde{r}} + \frac{2}{3 \tilde{r}^2} - \frac{\ln(1+\tilde{r})}{\tilde{r}^3}\right.\\ 
&\left.{} - \frac{1}{3 \tilde{r}^3 (\tilde{r} +1)} + \frac{1}{3 \tilde{r}^3} - \frac{1}{3 (1+\tilde{r})}  + \frac{\ln(1+\tilde{r})}{\tilde{r}}\right]\\ 
&+\eta \tilde{R}^2 \left[ \frac{1}{2 \tilde{r}^3}- \frac{2}{ \tilde{r}^4} + \frac{3 \ln(1+\tilde{r})}{\tilde{r}^5}+\right.\\ 
&\left.{}+\frac{1}{ \tilde{r}^5 (\tilde{r} +1)} - \frac{1}{\tilde{r}^5}\right]\quad\mbox{(prolate)}
\end{split}\end{equation}
(shown for $\eta=1/3$ in the left panel of Fig.\ \ref{fig.DMM}),
where $R=\sqrt{x^2+y^2}$ is the radius in the equatorial plane and $\tilde{R}=R/r_s$.
The oblate NFW model ($\epsilon=0$), maintaining now the names of the variables $x$, $y$ and $z$ as in Eq.s\ (\ref{eq.DC}) and (\ref{eq.PC}), has density distribution
\begin{equation}
\label{eq.DO}
\tilde{\rho}(R,z) = \frac{1}{\tilde{r}( 1+\tilde{r})^2} - \frac{\eta \tilde{z}^2}{\tilde{r}} \frac{1+3\tilde{r}}{\tilde{r}^2 (1+\tilde{r})^3}\quad\mbox{(oblate)},
\end{equation}
(shown for $\eta=1/3$ in the right panel of Fig.\ \ref{fig.DMM}) and gravitational potential 
\begin{equation}
\begin{split}
\label{eq.PO}
\tilde{\Phi}(R,z) =&  - \frac{\ln(1+\tilde{r})}{\tilde{r}} +\eta \left[-\frac{1}{6 \tilde{r}} + \frac{2}{3 \tilde{r}} - \frac{\ln(1+\tilde{r})}{\tilde{r}^3}+\right.\\  
&- \frac{1}{3 \tilde{r}^3 (1+\tilde{r})}+ \left.\frac{1}{3 \tilde{r}^3} - \frac{1}{3 (1+\tilde{r})}  + \frac{\ln(1+\tilde{r})}{\tilde{r}}\right]\\ 
&+\eta \tilde{z}^2 \left[ \frac{1}{2 \tilde{r}^3} - \frac{2}{ \tilde{r}^4}+ \frac{3 \ln(1+\tilde{r})}{\tilde{r}^5}+ \frac{1}{ \tilde{r}^5 (1+\tilde{r})}+\right.\\
&\left.{}- \frac{1}{\tilde{r}^5}\right]\quad\mbox{(oblate)}
\end{split}
\end{equation}
(shown for $\eta=1/3$ in the right panel of Fig.\ \ref{fig.DMM}).

In both cases, $z$ is the symmetry axis. 
Given that the first order terms of Eq.s (\ref{eq.DP}) and (\ref{eq.DO}) are $\propto \tilde R^2/\tilde r$ or $\propto \tilde z^2/\tilde r$, respectively, the subtraction of parent density is more significant in the outer regions. For $\eta\to 1/3$, it induces a peanutshaped distribution sufficiently far from the center (see Fig.\ \ref{fig.DMM}).

\begin{figure}
   \centering
   \includegraphics[width=0.5\textwidth]{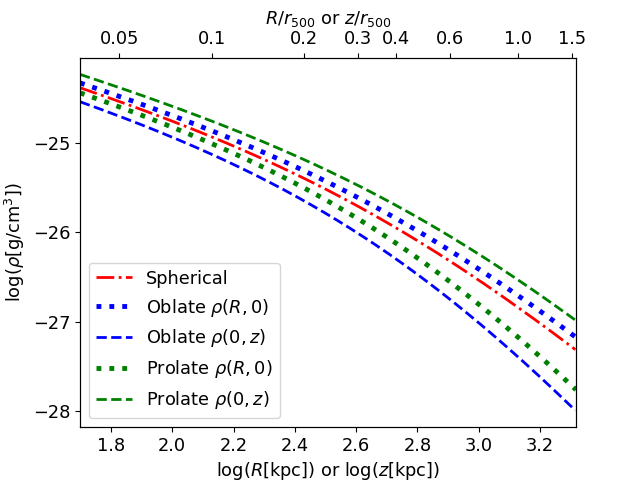}
   \includegraphics[width=0.5\textwidth]{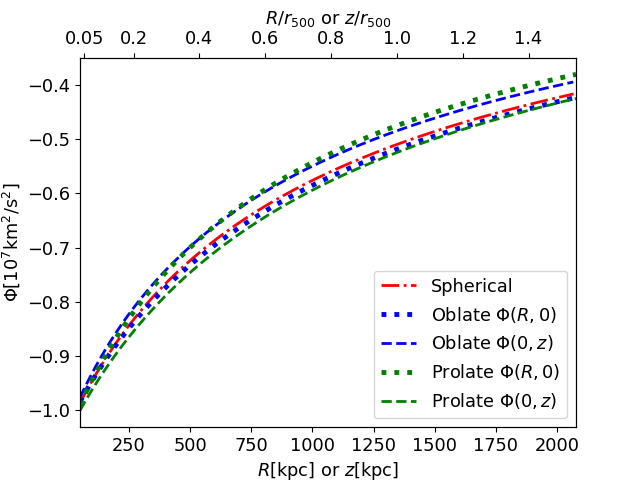}
   \caption{Profiles of density (upper panel) and gravitational potential (lower panel) of DMS (red lines), DMO (blue lines) and DMP (green lines) halo models (see Table \ref{tab.DM}).
   In particular, for our axisymmetric models we plot the density and gravitational potential profiles along the symmetry axis (dashed lines) and in the equatorial plane (dotted lines).
   The top axis in both panels indicates the distance from the center normalized to $\rcinque$ of DMS halo model 
   ($\rcinque=1345\, \mathrm{kpc}$; we note that the values of $\rcinque$ of our three halo models differ by less than 2\%).
   }
   
    \label{fig.DMH}%
    \end{figure} 

\subsection{Realistic halo models for massive clusters}
\label{sec.2.2}

A variety of halo shapes are expected from cosmological simulations (e.g.\ \citealt{Bett12,Henson17}; see also section 7.5.3 of \citealt{C19}), depending mainly on the halo merging history. When approximating the halos as ellipsoids, even if the majority of them is triaxial, the fact that the ratio of two of the three principal semiaxes is close to unity justifies the use of the spheroidal approximation for the description of these halos. However, for one of our models we adopt the spherical approximation, which is appropriate when the smallest-to-largest axial ratio is close to unity.

Using the density-potential pairs presented in Sect.\ \ref{sec.2.1}, we build our halo models as follows. The prolate and oblate NFW models (represented by Eq.s \ref{eq.DP}-\ref{eq.PP} and \ref{eq.DO}-\ref{eq.PO}, respectively, which both give for $\eta=0$ the spherical NFW model) are parameterized by $\rhon$, $\rs$ and $\eta$. 
To be as far as possible consistent with the predictions of cosmological simulations on the smallest-to-largest axial ratio \citep[see][]{A06}, in our spheroidal halo models, we assume the largest possible flattening ($\eta=1/3$) compatible with everywhere positive DM density distribution (see Sect.\ \ref{sec.2.1}).

When a spherical NFW model is considered in the cosmological context, the parameters $\rhon$ and $\rs$ can be expressed as functions of other two parameters, the virial mass $\MDelta$ and the concentration $\cDelta$, which are routinely measured in cosmological simulations (e.g. \citealt{A14}) and estimated for the halos of observed clusters of galaxies (e.g. \citealt{E10}). 
$\MDelta$ is the mass measured within a sphere of radius $\rDelta$, within which the average halo density is $\Delta \rhocrit(z)$, where the dimensionless quantity $\Delta$ is the overdensity and $\rhocrit(z)=3H^2(z)/(8 \pi G)$ is the critical density of the Universe at redshift $z$.  
The halo concentration is $\cDelta=\rDelta/\rminus$, where $\rminus$ is the radius where the logarithmic slope of the angle-averaged density profile is $-2$. 
For the spherical NFW model $\rs=\rminus=\rDelta/\cDelta$, where 
\begin{equation}
\label{eq.r_200}
\rDelta=\left[\frac{\MDelta}{(4/3)\pi \Delta \rhocrit(z)} \right]^{1/3},
\end{equation}
and we infer $\rhon$ from $\cDelta$ as
\begin{equation}
    \rhon=\frac{\Delta}{3} \frac{\rhocrit(z)\cDelta^3}{\ln(1+\cDelta)-{\cDelta}/(1+\cDelta)}.
\end{equation}

We now focus on the case of the standard overdensity value $\Delta=200$, and thus consider $\rdue$, $\Mdue$ and $\cdue=\rdue/\rminus$. 
To construct our specific spherical NFW, hereafter referred to as "dark matter spherical" (DMS) model, we set $\Mdue=10^{15}\,\mathrm{M_\odot}$ and $\cdue=3.98$, in agreement with the mass-concentration relation of \citet{A14} at redshift $z\approx 0$. 

For the spheroidal halo models, we first compute the mass within the sphere of radius $r$ 
\begin{equation}
\label{eq.M_sph}
M(r)=4\pi \int_{0}^{r} \left(\int_{0}^{\sqrt{r^2-z^2}} \rho(R,z) R d R\right) d z,
\end{equation}
where $\rho(R,z)$ is given by Eqs. (\ref{eq.DP}) or (\ref{eq.DO}) for the prolate and oblate NFW models, respectively. We then estimate $\rdue$ and $\rminus$ in the following way. The average density within the sphere of radius $r$ is $\langle \rho \rangle(r)=3M(r)/(4 \pi r^3)$;
while the angle-averaged density profile $\rhoshell(r)$ is estimated by measuring the average density within concentric spherical shells
\begin{equation}
\label{eq.rho_shell}
\rhoshell(r)=\frac{3[M(r+\delta r/2)-M(r-\delta r/2)]}{4 \pi [(r+\delta r/2)^3-(r-\delta r/2)^3]},
\end{equation}
where $\delta r=0.8\,\mathrm{kpc}$ is the thickness of the shell centered at the radius $r$. $\rDelta$ is thus defined to be such that $\langle \rho \rangle(\rDelta)\simeq \Delta\rhocrit(z)$,
and $\rminus$ to be such that 
\begin{equation}
\label{eq.r_-2}
\left[\frac{d\ln\rhoshell}{d\ln r}\right]_{r=r_{-2}}\simeq-2.
\end{equation}
The above equations can thus be used to estimate $\Mdue=M(\rdue)$ and $\cdue$ for our flattened halo models. 
In practice, to build the oblate and prolate NFW halo models, hereafter referred to as "dark matter oblate" (DMO) and "dark matter prolate" (DMP) models, respectively, we select pairs of values of $\rhon$ and $\rs$ such that $M_{200}\approx 10^{15}\,\mathrm{M_\odot}$ and $\cdue$ is consistent with the $z\approx0$ mass-concentration relation of \citet{A14}. 
The parameters of the halo models DMS, DMP and DMO are reported  in Tab. \ref{tab.DM}.  
The corresponding density and gravitational potential profiles along the symmetry axis and in the equatorial plane are shown in Fig. \ref{fig.DMH}.
The upper panel of Fig.\ \ref{fig.DMH} shows that, comparing models with approximately the same mass, because of the outward-increasing directional subtraction of parent density discussed in Sect.\ \ref{sec.2.1} (see Fig.\ \ref{fig.DMM}), the prolate model has $\rho(R,0)$ steeper and $\rho(0,z)$ shallower than the density profile of the spherical model, and vice versa for the oblate model.
Analogous (but weaker) trends are found in the gravitational potential profiles (lower panel of Fig.\ \ref{fig.DMH}).

\begin{table}
      \caption[]{Parameters of the adopted NFW halo models. 
      }
         \label{tab.DM}
     $$ 
         \begin{array}{p{0.1\linewidth}lllll}
            \hline
            \noalign{\smallskip}
            Model & \rhon[\mathrm{g/cm^3}] & \rs[\mathrm{kpc}] & \eta & \Mdue[\mathrm{M_\odot}] & \cdue\\
            \noalign{\smallskip}
            \hline
            \noalign{\smallskip}
            DMS & 4.8\times 10^{-26} & 519 & 0 & 1.00\times10^{15} & 3.98 \\
            DMO & 4.6\times 10^{-26} & 600 & 1/3 & 1.00\times10^{15} & 3.96 \\
            DMP & 4.8\times 10^{-26} & 700 & 1/3 & 1.01\times10^{15} & 4.27 \\
            \noalign{\smallskip}
            \hline
         \end{array}
     $$ 
     \tablefoot{We refer to these Navarro-Frenk-White models as "dark matter spherical" (DMS), "dark matter oblate" (DMO), and "dark matter prolate" (DMP) one, respectively.}
   \end{table}

\section{Building cool-core clusters models with rotating ICM}
\label{sec.3}

In this Section we present axisymmetric rotating models of the ICM that, in the absence of net cooling or heating, is in equilibrium in a given axisymmetric gravitational potential, representative of an isolated cluster. The ICM is sufficiently dense to cool in timescales much shorter than the Hubble time in the cluster core and thus to flow into the center of the gravitational potential well. 
However, as already mentioned in the Introduction, the effect of cooling is expected to be efficiently counteracted by heating  mechanisms, such as AGN and stellar feedback.
Thus, the adoption of stationary models of the ICM is justified as long as there is balance between cooling and heating in a time-averaged sense \citep[e.g.][]{MC12}, provided the cluster does not undergo major interactions.

\subsection{The equilibrium of rotating ICM in a cool-core cluster}
\label{sec.3.1}

Assuming that the total gravitational potential of the cluster $\Phi$ is time-independent and axisymmetric, we can build simple models of stationary rotating ICM by considering that the angular velocity of the gas is stratified over cylinders (and thus that the gas distribution is barotropic, i.e.\ with pressure stratified over density\footnote{More general (baroclinic) models, not explored in this work, have vertical gradients of angular velocity, and pressure not stratified over density.}).
Under these hypotheses, neglecting magnetic fields (which are dynamically unimportant for the ICM; see, e.g., \citealt{BRU13}),
the gas mass density $\rhogas(R,z)$ and pressure $p(R,z)$ are related by 
$\nabla p=-\rhogas \nabla\Phieff$, 
\begin{equation}
\label{eq.RE2}
\Phieff(R,z)=\Phi(R,z)-\Phi(\Rstar,\zstar)-\int_{\Rstar}^{R} \frac{u_{\phi}^2 (R')}{R'} dR',
\end{equation}
is the effective potential and $u_\phi(R)$ is the gas rotation velocity and $(\Rstar,\zstar)$ a reference point (e.g. \citealt{T78}).

From observations and hydrodynamical simulations, there is evidence that the ICM is well described by polytropic distributions, essentially independent of the halo mass \citep[e.g.][]{GPOLY}, in which the pressure is stratified over the density as a power law $p=\pstar (\rhogas/\rhogasstar)^{\gamma '}$,
where $\gamma'$ is the polytropic index, $\pstar=p(\Rstar,\zstar)$ and $\rhogasstar=\rhogas(\Rstar,\zstar)$.

Hereafter, we model the ICM in a cool-core cluster through a two-component composite polytropic distribution (e.g.\ \citealt{B13}), by assuming a polytropic index $\gammaout>1$ in the outer region and $\gammain<1$ in the cool core. It is convenient to adopt $(\Rstar,\zstar)=(\Rbreak,0)$, where $\Rbreak$ is a model parameter that defines the size of the cool core. For any outward-increasing axisymmetric potential, defining $\Delta\Phieff(R,z)=\Phieff(R,z)-\Phieff(\Rbreak,0)$,  we have $\Delta\Phieff(R,z)> 0$ in the outer region, and $\Delta\Phieff(R,z)\leq 0$ in the cool core. Assuming the ideal gas equation of state, the polytropic distributions of temperature and density of the ICM, in our models of cool-core clusters, are given by
\begin{equation}
\label{eq.CC1}
n(R,z)=\nstar \left[1- \frac{\gammaout-1}{\gammaout} \frac{\mu \proton}{\kboltz \Tstar} \Delta \Phieff(R,z)\right]^{\frac{1}{\gammaout-1}}
\end{equation}
and
\begin{equation}
\label{eq.CT1}
\Tgas(R,z)=\Tstar \left(\frac{n(R,z)}{\nstar} \right)^{\gammaout-1},
\end{equation}
where $\Delta\Phieff(R,z)>0$,
and  by
\begin{equation}
\label{eq.CC2}
n(R,z)=\nstar \left[1- \frac{\gammain-1}{\gammain} \frac{\mu \proton}{\kboltz \Tstar} \Delta \Phieff(R,z)\right]^{\frac{1}{\gammain-1}}
\end{equation}
and
\begin{equation}
\label{eq.CT2}
\Tgas(R,z)=\Tstar\left(\frac{n(R,z)}{\nstar} \right)^{\gammain-1},
\end{equation}
where $\Delta\Phieff(R,z)\leq0$. 
Here $n=\rhogas/(\mu \proton)$ is the gas number density, and $\nstar=\rhogasstar/(\mu \proton)$; $\mu$, $\proton$, and $\kboltz$ are the mean molecular weight (taken equal to 0.6), the proton mass and the Boltzmann constant, respectively.

\subsection{Rotation law and effective potential}

Though the ICM rotation velocity curve is poorly constrained observationally \citep[see][for an attempt]{L19}, it is reasonable to expect that it could have a relatively steep rise of azimuthal velocity in the cluster center, a peak at intermediate radii, and a gradual fall in the outskirts \citep[see][]{BALDI17,A23}.
In particular, following \cite{B13}, we adopt the rotation law
\begin{equation}
\label{eq.RE4}
u_{\phi}(R)=\uzero \frac{S}{(1+S)^2},
\end{equation}
where $S\equiv R/\Rzero$, $\Rzero$ is a reference radius and $\uzero$ a reference speed. 

Substituting the rotation law (\ref{eq.RE4}) in Eq. (\ref{eq.RE2}), and integrating the rotational component of the effective potential, we get the analytic effective potential associated with this rotation law
\begin{equation}
\label{eq.RE5}
\Phieff(R,z)= \Phi(R,z)-\Phi(\Rstar,\zstar) -\left[I(R)-I(\Rstar)\right],
\end{equation}
where
\begin{equation}
I(R')=\uzero^2\left[\frac{1}{3}\left(1+\frac{R'}{\Rzero}\right)^{-3}-\frac{1}{2}\left(1+\frac{R'}{\Rzero}\right)^{-2}\right].
\label{eq:IR}
\end{equation}

\label{sec.3.2}
\begin{figure}
   \centering
   \includegraphics[width=0.5\textwidth]{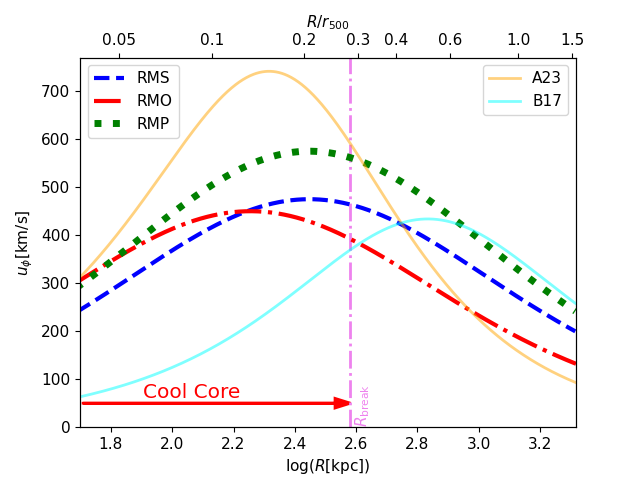}
   \caption{Profiles of ICM rotation speed of our cluster models with spherical (RMS; blue dashed curve), oblate (RMO; red dotted-dashed curve), and prolate (RMP; green dotted curve) halos. For comparison, we also show as solid curves the average rotation speed profiles of the ICM found in clusters formed in cosmological simulations: in particular, the cyan and orange curves are obtained from the functional forms given by, respectively,  \citet{BALDI17} and \citet{A23}, assuming $\rDelta$ and $\MDelta$ as in model DMS. 
   In the top axis, the radial coordinate in the meridional plane is normalized to $\rcinque$ of RMS model.    
   The vertical line indicates $R=\Rbreak$ for the spherical model, approximately enclosing the cool-core region (red arrow), which has roughly the same extent in all models (see Tab.\ \ref{tab.ROT}).}
    \label{fig.RC}
    \end{figure}

\begin{table*}
      \caption[]{Parameters of the cluster models with rotating ICM.}
         \label{tab.ROT}
     $$ 
         \begin{array}{p{0.1\linewidth}llllllll}
            \hline
            \noalign{\smallskip}
            Model & {\rm Halo} & \Rbreak[\mathrm{kpc}] & \neestar[\mathrm{cm^{-3}}] & \Tstar[\mathrm{keV}] & \uzero[\mathrm{km/s}] & \Rzero[\mathrm{kpc}] & \gammain & \gammaout\\
            \noalign{\smallskip}
            \hline
            \noalign{\smallskip}
            RMS & {\rm DMS} & 380 & 2.5\times 10^{-3} & 7.3 & 1900 & 280 & 0.83 & 1.19\\
            RMO & {\rm DMO} & 420 & 2.2\times 10^{-3} & 7.4 & 1800 & 180 & 0.82 & 1.20\\
            RMP & {\rm DMP} & 360 & 2.5\times 10^{-3} & 7.4 & 2300 & 280 & 0.85 & 1.19\\
            \noalign{\smallskip}
            \hline
         \end{array}
     $$ 
     \tablefoot{We refer to these cluster models with rotating ICM as "rotating model spherical" (RMS), "rotating model oblate" (RMO) and "rotating model prolate" (RMP) depending on the spherical, oblate and prolate DM halo models assumed, respectively. The corresponding halo models DMS, DMP, and DMO (cited in the second column) are defined in Sect.\ \ref{sec.2.2}.}
   \end{table*}

\subsection{Three representative models of massive cool-core clusters with rotating ICM}
\label{sec.ROT}
Without focusing on a particular cluster, we propose three models with rotating ICM representative of the observed population of massive ($\Mdue \approx 10^{15}{\rm M}_\odot$) cool-core clusters, dubbed "rotating model spherical" (RMS), "rotating model oblate" (RMO) and "rotating model prolate" (RMP).
In all these models we assume that the gas follows a two-component composite polytropic distribution described by Eqs.\ (\ref{eq.CC1}-\ref{eq.CT2}), and that the rotation law has the functional form (\ref{eq.RE4}). 
The effective potential is thus in the form of 
Eqs.~(\ref{eq.RE5}-\ref{eq:IR}). 
In all cases, to compute the intrinsic and emission properties of the ICM, we assume metallicity $Z=0.3\,\mathrm{Z_{\odot}}$ (where $\mathrm{Z_{\odot}}$ is the solar metallicity reported in \citealt{A89}), implying $n/\nee=1.94$, where  $n=\nii+\nee$ is the gas number density, $\nee$ is the electron number density and $\nii$ is the ion number density (assuming full ionization).

In model RMS the total gravitational potential $\Phi$ is given by the spherical gravitational potential of the halo model DMS described in Sect. \ref{sec.2.2}. 
In models RMO and RMP, the total gravitational potential is axisymmetric, being, respectively, the potential of the oblate halo model DMO and of the prolate halo model DMP, described in \ref{sec.2.2}. 
The values of the plasma parameters  $\Rbreak$, $\neestar$, $\Tstar$, $\gammain$ and $\gammaout$, and of the parameters of the rotation pattern $\Rzero$ and $\uzero$ are reported for all the models in Tab. \ref{tab.ROT}.
The ICM rotation speed profiles of the three models, with peak rotation speeds in the range  400-$600\,\mathrm{km/s}$,  are shown in Fig. \ref{fig.RC}. 
In the same figure we plot, for comparison, the average rotation speed profiles of clusters formed in MUSIC\footnote{The synthetic clusters of \cite{BALDI17} are selected from the MUSIC-2 sample \citep{Sembolini13} having $M_{200}>5\times 10^{14} h^{-1} \,\mathrm{M_\odot}$, where $h=H_0/(100 \,\mathrm{km/s/Mpc})$.
The corresponding curve in Fig.\ \ref{fig.RC} is built using data taken from table 4 of \cite{BALDI17}, for gas-VP2b rotation curve in the so-called AGN simulation.} \citep{BALDI17} and MACSIS\footnote{The MACSIS cluster sample \citep{Barnes17} have friends-of-friends masses at redshift $z=0$ $\gtrsim 10^{15} \,\mathrm{M_\odot}$. The corresponding curve in Fig.\ \ref{fig.RC} is built using data taken from table B2 of \cite{A23} for the $\Mcinque<9.7\times 10^{14}\,\mathrm{M_\odot}$ subsample in the so-called gas-aligned case.} \citep{A23} cosmological simulations. 
Our rotation speed profiles are in between the average profiles found by \cite{BALDI17} and  \cite{A23}, and can thus be considered, in this sense, cosmologically motivated. 
Moreover, in Sect.\ \ref{sec:comparison} we show that our three rotating models are realistic, in the sense that they have properties consistent with the currently available observational data of real massive clusters.

\begin{figure*}
   \centering
   \includegraphics[width=0.49\textwidth]{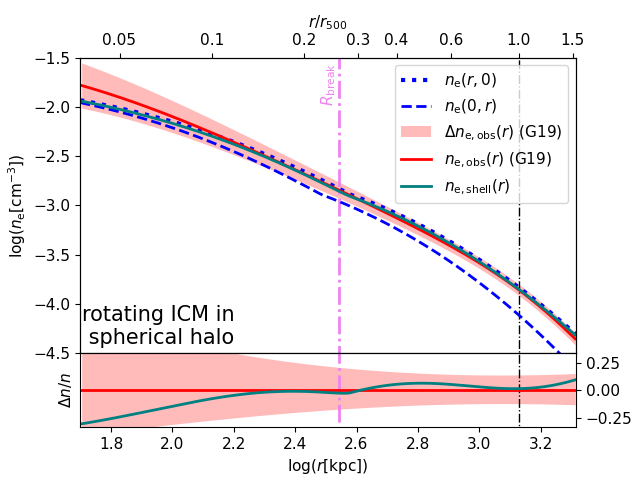}
   \includegraphics[width=0.49\textwidth]{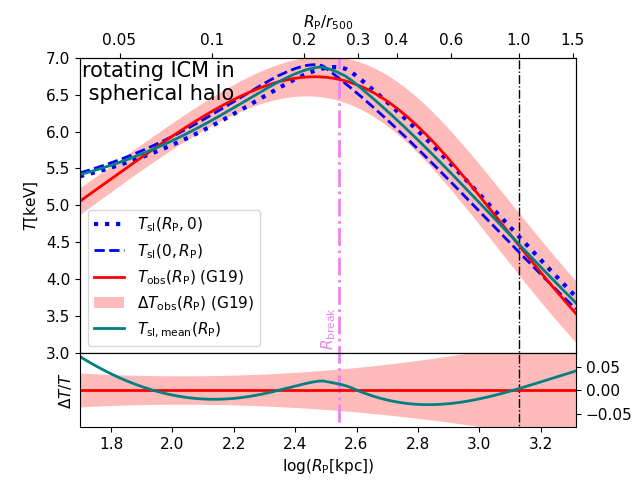}
   \caption{Thermodynamic profiles of the ICM in "rotating model spherical" (RMS) model. Upper panels. Radial (dotted) and vertical (dashed) profiles of electron number density (left panel) and spectroscopic-like temperature (right panel) for model RMS (blue lines) compared with the corresponding average observed profiles (red solid lines) and their scatter (red shaded band), taken from \citet[][G19 in the legends]{G19}. Here, $\nshell$ (green solid line; left panel) is the angle-averaged (see Sect. \ref{sec.2.2}) density profile of model RMS, and $\Tmean(R_{\rm P})=\left[\Tsl(R_{\rm P},0)+\Tsl(0,R_{\rm P})\right]/2$ (green solid line; right panel) is its average spectroscopic-like temperature profile.
   Lower panels. Departure of average profile (green solid lines; see above) of density (left panel) and spectroscopic-like temperature (right panel) of model RMS from the average observed profiles (red solid lines) with their scatter (red shaded band).
   The spherical radius $r$, and the radius in the plane of the sky $R_{\rm P}=\sqrt{x^2+z^2}$ are given in kpc in the bottom axis and normalized to $\rcinque\simeq 1345 \,\mathrm{kpc}$ in the top axis. The violet and black dot-dashed vertical lines indicate $\Rbreak$ and $\rcinque$, respectively.
   The virial temperature of this model, defined as in equation 59 of \cite{V05}, is $T_{200}\simeq 6.46$ keV.
   }
    \label{fig.RMS}
    \end{figure*}
\begin{figure*}
   \centering
   \includegraphics[width=0.49\textwidth]{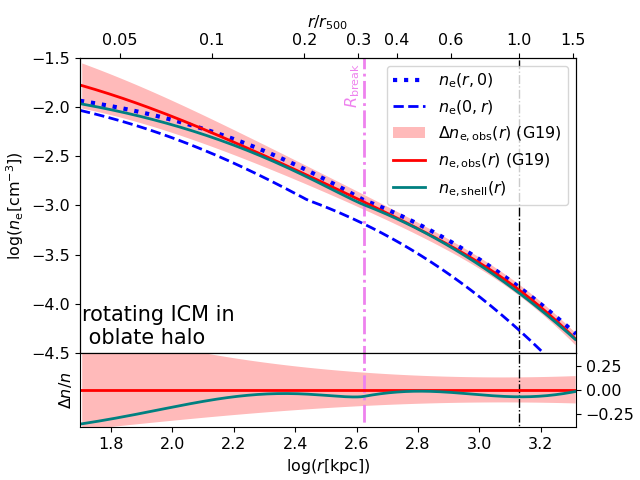}
   \includegraphics[width=0.49\textwidth]{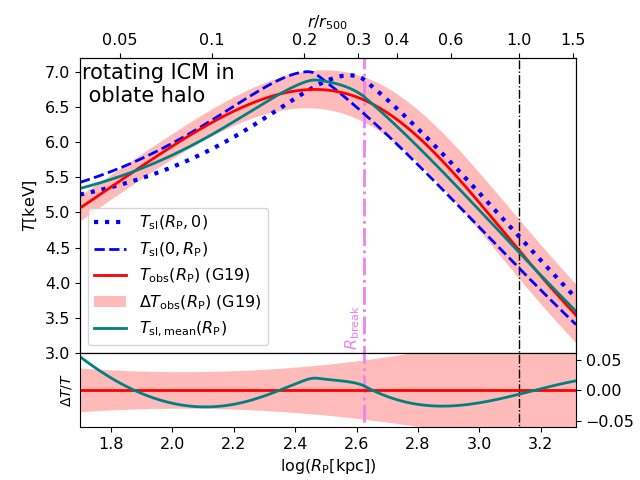}
   \caption{Density (left panel) and spectroscopic-like temperature (right panel) profiles of "rotating model oblate" (RMO) model. The figure display is the same as Fig. \ref{fig.RMS}, but for RMO model $r_{500}\simeq 1346 \,\mathrm{kpc}$ and $T_{200}\simeq 6.45$ keV.}
    \label{fig.RMO}
    \end{figure*}
\begin{figure*}
   \centering
   \includegraphics[width=0.49\textwidth]{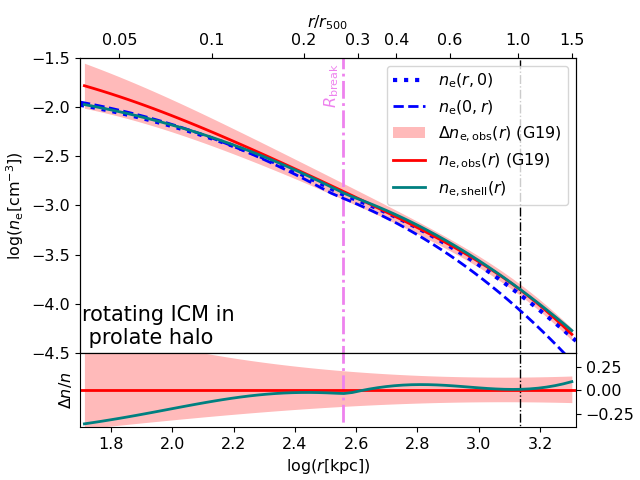}
   \includegraphics[width=0.49\textwidth]{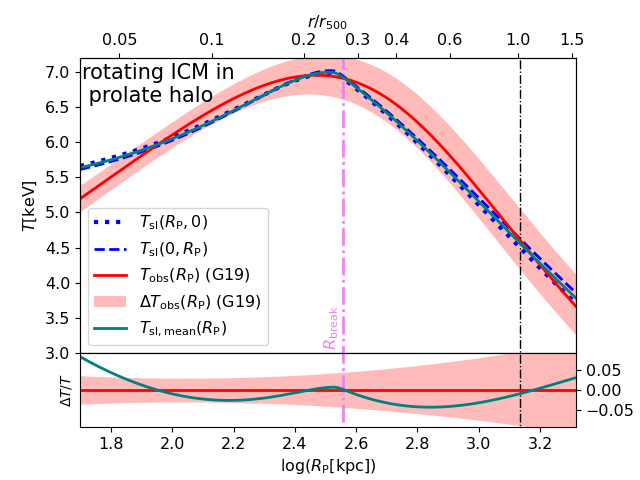}
   \caption{Density (left panel) and spectroscopic-like temperature (right panel) profiles of "rotating model prolate" (RMP) model. The figure display is the same as Fig. \ref{fig.RMS}, but for RMP model $r_{500}\simeq 1366 \,\mathrm{kpc}$ and $T_{200}\simeq6.53$ keV.}
    \label{fig.RMP}
    \end{figure*}

\section{Comparison with observations}
\label{sec:comparison}

Here we compare with observational data some properties of the cool-core cluster models with rotating ICM presented in Sect.\ \ref{sec.ROT}.

\subsection{Thermodynamic profiles of the ICM}
\label{sec.THERMO}

Two directly observable quantities of the ICM are the emission measure, which is a proxy for gas density, and the spectroscopic temperature ($\Tsp$), which is the temperature associated with the emission in the X-ray spectrum. 
Despite the difficulty to find an analytic expression of the spectroscopic temperature, \cite{M04} have found a good approximation of it, called the spectroscopic-like temperature, which, 
for an axisymmetric cluster with symmetry axis $z$ orthogonal to the line of sight, is given by 
\begin{equation}
    \label{eq.Tsl}
    \Tsl(x,z)=\left.\int_{|x|}^{\infty} \frac{\nii \nee \Tgas^{1/4} \hat{r} d\hat{r}}{\sqrt{\hat{r}^2 - x^2}}\middle/\int_{|x|}^{\infty} \frac{\nii \nee \Tgas^{-3/4} \hat{r} d\hat{r}}{\sqrt{\hat{r}^2 - x^2}}\right.,
\end{equation}
where $\Tgas$ is the gas temperature (in this work, given by Eq.s \ref{eq.CT1}, \ref{eq.CT2}) and $\hat r$ is the radius in the plane at height $z$, parallel to the equatorial plane. 
Here, $x$ and $z$ are the coordinates in the plane of the sky, with the origin in the cluster center.

According to the cosmological framework of formation and evolution of cosmic structures, the population of galaxy clusters is expected to be homogeneous, with ``universal'' profiles of the thermodynamic quantities (density, temperature, pressure, and entropy) of the ICM that depend only on the mass and redshift of the halo \cite[see e.g.][]{V06,Pratt10,Arnaud10,Eckert12,G19,ettori23}.
This is particularly true in the regions dominated by the action of gravity.

Recently, the combination of high-quality data of thermal Sunyaev-Zeldovich effect (\citealt{S72}) and of X-ray observations have allowed \cite{G19} to reconstruct the universal thermodynamic profiles of the XMM Cluster Outskirts Project (X-COP) sample (\citealt{X-COP}) out to $\rdue$ with an unprecedented accuracy\footnote{The X-COP sample consists of 13 nearby, massive galaxy clusters selected on the basis of signal-to-noise ratio of the Sunyaev-Zeldovich effect as resolved in the \emph{Planck} maps \citep{P14}. Five of these objects are classified as relaxed, cool-core systems accordingly to their central entropy.}
(see also \citealt{V06} and \citealt{NV07} for the discussion on the reliability of the reconstruction method). 

We thus compare our models of the rotating ICM in equilibrium in cool-core clusters of $\Mdue\approx 10^{15}\,\mathrm{M_\odot}$ with these thermodynamic profiles in Fig.s \ref{fig.RMS}-\ref{fig.RMP}, where the observed temperature is $\Tsp$.
We note that in the inner regions of the cool core (i.e. $r<60\,\mathrm{kpc}$) the spectroscopic-like temperature of the models departs significantly from the observed profile of the spectroscopic temperature, this discrepancy is not very meaningful, given the observational limitations on the recovery of the thermodynamic properties in such central regions. 
The thermodynamic properties of models RMS, RMO and RMP, with different halo shapes and rotation patterns, are thus reasonably representative of the average properties of the ICM in massive cool-core clusters.

Once shown that the ICM pressure is stratified over the ICM density following a power-law function (i.e.\ that the distribution is polytropic), in the X-COP sample \cite{GPOLY} have found polytropic indices that, depending on the cluster radius, span from 0.75 (in the inner region) to 1.25 (in the outer region), independent of the cluster mass. The polytropic indices of our rotating ICM models (RMS, RMO and RMP), $\gammain \simeq0.8$ and $\gammaout \simeq1.2$ (see Tab. \ref{tab.ROT}), are fully consistent with those of the observed clusters. 

We note that reproducing the observed thermodynamic profiles under the assumption of rotating ICM is not guaranteed: this is discussed in Appendix \ref{A.THERMO}, where we present an illustrative example of a model with strongly rotating ICM, which fails to reproduce some characteristic features of the observed population of massive clusters.

\subsection{Flattening of the X-ray surface-brightness distributions}
\label{sec.ELL}

The gas rotation and halo flattening leave a trace in the shape of the X-ray surface-brightness distribution. Here, we compare the shape of the X-ray surface-brightness distribution in our models and in real massive clusters.  
One way to account for the departure of the iso-surface brightness contours from the circular shape is through an average axial ratio, based on the inertia's tensor of surface brightness distribution \citep[see][]{B92,B94}.
 
Assuming to observe  our models edge-on (i.e.\ with symmetry axis orthogonal to the line of sight), the surface brightness is
\begin{equation}
\label{eq.OBS3}
\Sigma(x,z)=2 \int_{|x|}^{\infty} \frac{\nii \nee \Lambda (T) \hat{r} d\hat{r}}{\sqrt{\hat{r}^2 - x^2}},
\end{equation}
where $\Lambda(T)$ is the cooling function (in particular we take $\Lambda$ from \citealt{T01}, for $Z=0.3\,\mathrm{Z_\odot}$). 
\begin{figure*}
   \centering
   \includegraphics[width=0.49\textwidth]{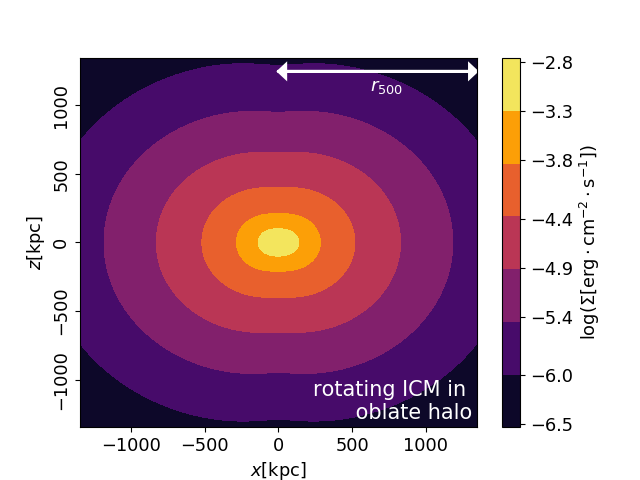}
   \includegraphics[width=0.49\textwidth]{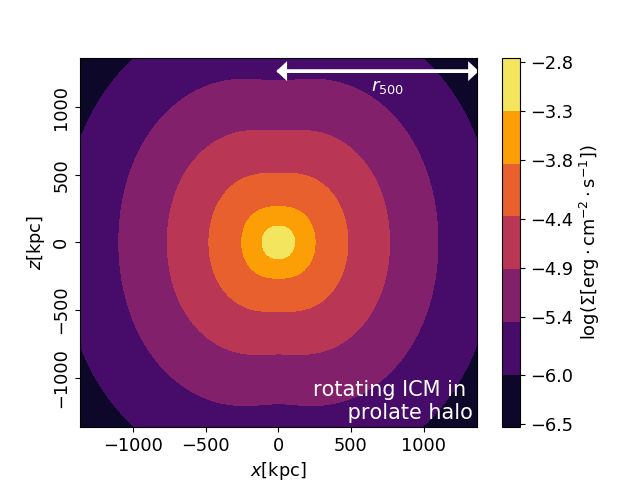}
   \caption{Surface-brightness maps of models RMO (left panel) and RMP (right panel). 
    The boxes (with origin in the cluster center) extend out to $\approx \rcinque$ (see white arrows).}
    \label{fig.BRIGHT}
    \end{figure*}
Using Eq.\ (\ref{eq.OBS3}), we compute the surface-brightness distribution of our models, which is shown in Fig.\ \ref{fig.BRIGHT} for models RMO and RMP.

Given that the inertia's tensor of the surface brightness distribution is in diagonal form for a cluster observed edge-on, its diagonal terms are $I_{20}=\sum_{i=1}^P \Sigma_i x_i^2$ and $I_{02}=\sum_{i=1}^P \Sigma_i z_i^2$,
where $\Sigma_i$ is the surface brightness (given by Eq. \ref{eq.OBS3}) at the grid point of plane-of-the-sky coordinates $(x_i,z_i)$, called hereafter pixel, and $P$ is the total number of pixels. From the definition of diagonal terms, it follows that the average axial ratio is
$\zeta={I_{\min}}/{I_{\max}}$, where $I_{\max}=\max\{I_{20},I_{02}\}$ and $I_{\min}=\min\{I_{20},I_{02}\}$.

    \begin{figure}
   \centering
   \includegraphics[width=0.49\textwidth]{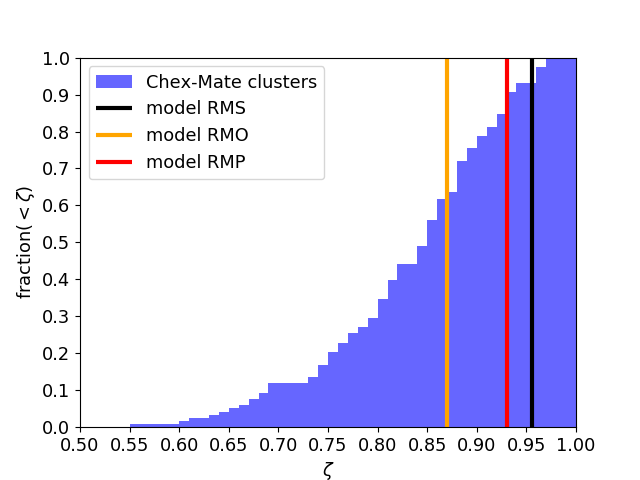}
   \caption{Comparison of the average axial ratio of surface-brightness distribution of models (vertical lines) RMS (black), RMO (yellow) and RMP (red) with the cumulative distribution of the average axial ratios measured for the CHEX-MATE clusters (blue) by \cite{C21}.}
    \label{fig.AXIAL}
    \end{figure}
In this work, we compare our models to the results obtained for the XMM Cluster Heritage Project (CHEX-MATE) sample\footnote{The CHEX-MATE sample is a signal-to-noise limited sample of 118 galaxy clusters detected by Planck via their Sunyaev-Zel’dovich effect; it is composed by two subsamples: the Tier-1, including the population of clusters at the most recent time ($z < 0.2$) and the Tier-2, with the most massive objects to have formed thus far in the history of the Universe; see \href{http://xmm-heritage.oas.inaf.it/}{http://xmm-heritage.oas.inaf.it/} for further details.} (\citealt{CHEX}), which contains both cool-core and non-cool-core clusters observed within their $\rcinque$. 
To match the clusters of this sample, we compute the average axial ratio of our cluster models only in the plane-of-the-sky region defined by $ \Rbreak \leq\vert x\vert\leq \rcinque$ and $\Rbreak \leq |z|\leq \rcinque$. 
In Fig.\ \ref{fig.AXIAL} we present the cumulative distribution of average axial ratios of CHEX-MATE clusters, where the 25th, 50th, and 75th percentiles are $\zeta=0.77$, $\zeta=0.85$, and $\zeta=0.89$, respectively (see also figure B.1 of \citealt{C21}).

The models RMS, RMO and RMP have,  respectively,  $\zeta=0.96$, $\zeta=0.87$, and $\zeta=0.93$, corresponding to the 93th, 62th and 85th percentiles of the distribution of the CHEX-MATE sample and thus are consistent with the less flattened population of massive clusters. 
The halos formed in cosmological simulations (having average ellipticity $\approx 0.5$; e.g. \citealt{A06}) tend to be more flattened than our aspherical halo models (having ellipticity $\sim 0.3$). The relatively high values of $\zeta$ of our cluster models are a consequence of the method adopted to build the density-potential pairs of our oblate and prolate halo models: given the requirement of everywhere positive halo density, the \cite{C05} method prevents from building highly flattened halos (see Sect.\ \ref{sec.2}). However, the flattening of our ICM models is due only to rotation and halo shape, 
while mergers, substructures, and anisotropic turbulence, all neglected in our models, are likely present in real clusters, where they can contribute to lower $\zeta$.

\begin{figure}
   \includegraphics[width=0.49\textwidth]{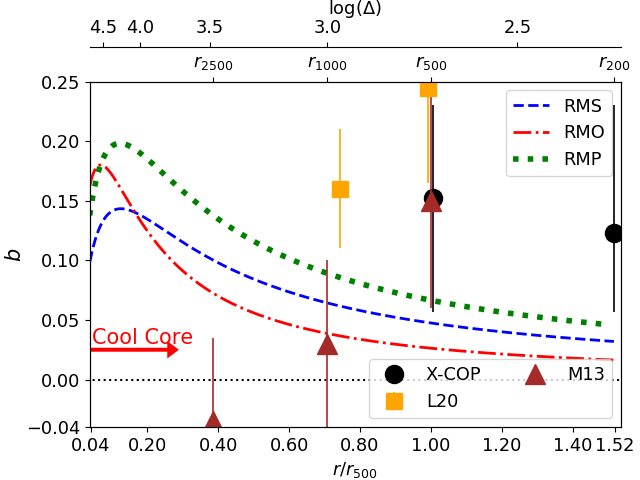}
   \caption{Profiles of the hydrostatic mass bias of models RMS (blue dashed line), RMO (red dashed-dotted line) and RMP (green dotted line), compared to the observational measurements (points): in particular, to average hydrostatic mass biases, taken from figure 5 of \citet[][black points]{E19}, from tables 2 and 4 of \citet[][orange points]{lovisari20}, and from table 4 of \citet[][brown points]{MAHDAVI2013}.
   The vertical error bars of observational data indicate 1$\sigma$ uncertainty on $b$. In the bottom axis, the radius is normalized to $\rcinque$ of the corresponding model, with the RMS model radii $r_{2500}$, $r_{1000}$, $r_{500}$, and $r_{200}$ reported on the plot top.
   The top axis reports the corresponding overdensity $\Delta$ of the RMS model  (for models RMO and RMP the overdensity values are very similar to those of model RMS).
    The radius corresponding to a given overdensity is computed from the true mass profile, which, in the case of observational data, is assumed to be obtained from the weak gravitational lensing analysis. The red arrow indicates the extent of the cool core.}
    \label{fig.b_all}
    \end{figure}

\subsection{Hydrostatic mass bias}
\label{sec:bias}
The mass recovered under the assumption of hydrostatic equilibrium and spherically symmetric gravitational potential is \citep[e.g.][]{lau13}
\begin{equation}
    \label{eq.MHE}
    \MHE(<r)=-\frac{r^2}{G\rhogasshell}\frac{\partial \pshell}{\partial r},
\end{equation}
where $\pshell$ and $\rhogasshell$ are, respectively, the angle-averaged (see Sect. \ref{sec.2.2}) pressure and density profiles. 
The hydrostatic mass bias profile is
\begin{equation}
    \label{eq.BHE}
    b(r)=1-\frac{\MHE(<r)}{\Mtrue(<r)},
\end{equation}
where $\Mtrue$ is the angle-averaged mass (Eq.\ \ref{eq.M_sph}) of the halo model that generates the gravitational potential, into which the ICM is in equilibrium.
Using  Eqs. (\ref{eq.MHE}) and (\ref{eq.BHE}), we compute $b(r)$ for our cluster models, which we plot in Fig. \ref{fig.b_all}, finding in all cases that the hydrostatic mass bias, except for the central region, tends to decrease with radius. 

The mass estimates from weak gravitational lensing are believed to be significantly less biased than those from X-ray observations \citep[e.g.][]{meneghetti10,Lee2018}, at least for nonmerging clusters \citep{lee23}. Thus, when we consider the hydrostatic mass bias of real clusters, we take the cluster mass from weak lensing as an estimate of $\Mtrue$.
In particular, in Fig. \ref{fig.b_all} we compare the hydrostatic mass bias of our cluster models to the following measurements:
\begin{itemize}
    \item The error-weighted average of the hydrostatic mass biases of the massive clusters in the X-COP sample, which are classified as relaxed, at true (i.e.\ obtained  from weak lensing measurements) $\rcinque$ and $\rdue$. The hydrostatic and weak lensing masses are determined by \cite{E19} and \cite{Herbonnet20}, respectively.
    \item The average hydrostatic mass bias of a large subsample of the \emph{Planck} Sunyaev-Zeldovich effect galaxy clusters (62 clusters of true masses in the range $3\times 10^{14}-2\times 10^{15} \,\mathrm{M_\odot}$, at $z<0.5$) at $1\,\mathrm{Mpc}$ and true $\rcinque$. The hydrostatic masses are determined by \cite{lovisari20}, while the weak lensing masses are taken from \cite{CatalogeWL15}.
    \item The average hydrostatic mass bias of the relaxed cluster subsample (most of which are found to have prominent cool cores) of the Canadian Cluster Comparison Project (50 clusters at $0.15<z<0.55$, selected with the X-ray spectroscopic temperature $>3\,\mathrm{keV}$), at true $r_{2500}$, $r_{1000}$ and $\rcinque$. The hydrostatic and weak lensing masses are determined by \cite{MAHDAVI2013} and \cite{Hoekstra12}, respectively.
\end{itemize}

As shown by Fig. \ref{fig.b_all}, the rotation support assumed in our cluster models is realistic, in the sense that it induces hydrostatic mass bias comparable to or lower than those detected in real clusters (with the exception of the estimate of \citealt{MAHDAVI2013} at $r_{2500}$; see Sect. \ref{sec.DISC} for a discussion).
On the basis of the comparison of the thermodynamic profiles of the ICM, shape of surface-brightness distribution, and hydrostatic mass bias of our cluster models with observations, we conclude that our models are consistent with the main cluster observables that are currently able to constrain the rotation speed of the ICM in cool-core clusters.

\section{Measuring rotation with X-ray spectroscopy}
\label{sec.DETECT} 

In the near future, the advent of the microcalorimeters, soft X-ray spectrometers such as \emph{Resolve} onboard \emph{XRISM}, a JAXA/NASA collaborative mission with ESA participation, will provide us with X-ray spectra at high spectral resolution \citep{Tashiro18}, allowing us to measure the line-of-sight component of the ICM velocity (e.g.\ \citealt{OTA18}) and thus estimate its rotation support.  
In this Section, using the configurations for \emph{Resolve}, we present a set of mock X-ray spectra of the rotating ICM in our cluster models and we assess the detectability of rotation with X-ray spectroscopy.

\begin{table}
      \caption[]{Characteristics of the mock pointings. }
         \label{tab.REG}
     $$ 
         \begin{array}{p{0.25\linewidth}lll}
            \hline
            \noalign{\smallskip}
            Region & \left\vert x \right \vert\,[\mathrm{kpc}] & z\,[\mathrm{kpc}] & Radius\,[\mathrm{kpc}]\\
            \noalign{\smallskip}
            \hline
            \noalign{\smallskip}
            R1 & 200 & 0 & 100  \\
            R2 & 650 & 0 & 150 \\
            R3 & 1150 & 0 & 250 \\
            \noalign{\smallskip}
            \hline
         \end{array}
     $$ 
     \tablefoot{Coordinates (second and third columns) and radius (fourth column) of circular regions of the mock observations in the plane of the sky (with the origin in the cluster center).}
   \end{table}

\subsection{Building mock spectra of the rotating ICM}
\label{sec.MOCK}

Here, we present our mock spectra, focusing primarily on the kinematic signatures. 
Given that, for a temperature of the ICM higher than $3\, \mathrm{keV}$, a mock multitemperature source spectrum (i.e.\ constructed from a multitemperature model) and the best fit to this spectrum with a singletemperature model are indistinguishable in the X-rays (\citealt{M04}), we directly simulate the X-ray thermal emission of the ICM of our models through a singletemperature model. 
In particular, we use the velocity Broadened Astrophysical Plasma Emission Code\footnote{\href{https://heasarc.gsfc.nasa.gov/xanadu/xspec/manual/node136.html}{https://heasarc.gsfc.nasa.gov/xanadu/xspec/manual/node136.html}.} (BAPEC), 
where a parameter accounts for a general broadening of the X-ray emission lines, including the thermal broadening of the ionized metals, and any other contribution in the form of ``Doppler broadening'' due to the cumulative effect of the different Doppler shifts caused by a distribution of the velocities of the ions. 
With this model, the Doppler shift of the lines is parametrized by an effective redshift ($\zeff$), which can be different from the cluster's redshift $z_0$ due to the action of a coherent, bulk motion, and their equivalent width is regulated by the metallicity which we fix to $0.3\,\mathrm{Z_{\odot}}$.
We observe our models of cool-core clusters edge-on, to maximize the contribution of rotation to the l.o.s.\ velocity, which is thus 
\begin{equation}
    \label{eq.vlos}
    \vlos(x,z)=\frac{1}{\Sigma(x,z)}\int_{\vert x \vert}^{\infty} \frac{\nii \nee \Lambda(T) \vert x\vert u_\phi(\hat r) d\hat r}{\sqrt{\hat{r}^2 - x^2}},
\end{equation} 
where $\Sigma(x,z)$ is given by Eq. (\ref{eq.OBS3}), $u_\phi(\vert x\vert)$ by (\ref{eq.RE4}) with parameters $\uzero$ and $\Rzero$ reported in Tab. \ref{tab.ROT}. 
To decouple the rotation from the contributions to the broadening of X-ray emitting lines, we observe sufficiently large regions, to be spatially resolved by the spectrometer \emph{Resolve}, where the ICM is either approaching or receding: in particular, we simulate the observation of regions R1, R2, and R3, reported in Tab. \ref{tab.REG}. 
The l.o.s.\ speed of the ICM in our cluster models is consistent with the observational upper limit on the rotation speed of $500\,\mathrm{km/s}$ in the cool cores of real galaxy clusters \citep[e.g.][]{S11,P15,BAMBIC}: for all models, $\vert \vlos\vert \lesssim 450\,\mathrm{km/s}$ in region R1, which belongs to the inner region. 
We find that the energy shift of a $6\,\mathrm{keV}$ line due to the rotation speed of $400 \,\mathrm{km/s}$ is $8 \,\mathrm{eV}$.
\emph{Resolve}, thanks to its energy resolution of $\sim 7 \,\mathrm{eV}$ at $E=6\,\mathrm{keV}$\footnote{\url{https://xrism.isas.jaxa.jp/research/analysis/manuals/xrqr_v2.1.pdf}}, has the potential to detect such an energy shift, unlike the currently available X-ray CCD detectors with energy resolution in the order of $\approx 100 \,\mathrm{eV}$.
Assuming $\vlos$ positive for approaching ICM, we compute $\zeff$ as
\begin{equation}
    \label{eq.zeff}
    \zeff=(1+z_0)\sqrt{\frac{1+\frac{\langle \vlos\rangle}{c}}{1-\frac{\langle \vlos\rangle}{c}}}-1,
\end{equation}
where we always take $z_0=0.05$. 
In this Section, $\langle...\rangle$ refers to the average along all the lines of sights, which cross one of the regions of Tab. \ref{tab.REG}: following \cite{R18}, we use as a weight for the average along the l.o.s.\ $\nii \nee\Lambda(T)$, except for the spectroscopic-like temperature, which is defined by Eq.\ (\ref{eq.Tsl}). 

\begin{table}
      \caption[]{Input parameters of our mock spectra.}
         \label{tab.PARA}
     $$ 
         \begin{array}{p{0.3\linewidth}llll}
            \hline
            \noalign{\smallskip}
             Model - Region & T [\mathrm{keV}] & norm & \zeff(R) &  \zeff(B) \\
            \noalign{\smallskip}
            \hline
            \noalign{\smallskip}
             RMS - R1 & 6.36 & 0.0072 & 0.0512 & 0.0488 \\
             RMS - R2 & 6.01 & 0.0014 & 0.0511 & 0.0489 \\
             RMS - R3 & 4.96 & 0.0007 & 0.0509 & 0.0491 \\
             RMO - R1 & 6.23 & 0.0078 & 0.0512 & 0.0488 \\
             RMO - R2 & 6.18 & 0.0014 & 0.0509 & 0.0491 \\
             RMO - R3 & 5.07 & 0.0007 & 0.0506 & 0.0494 \\
             RMP - R1 & 6.55 & 0.0061 & 0.0515 & 0.0485 \\
             RMP - R2 & 6.00 & 0.0012 & 0.0514 & 0.0486 \\
             RMP - R3 & 4.92 & 0.0005 & 0.0511 & 0.0489 \\
            \noalign{\smallskip}
            \hline
         \end{array}
     $$ 
     \tablefoot{The parameter $norm$ accounts for the normalization of the spectrum. We quote the effective redshift ($\zeff$) both for receding (identified by R) and approaching (by B) ICM.}
   \end{table}

\begin{figure*}
   \centering
   \includegraphics[width=0.49\textwidth]{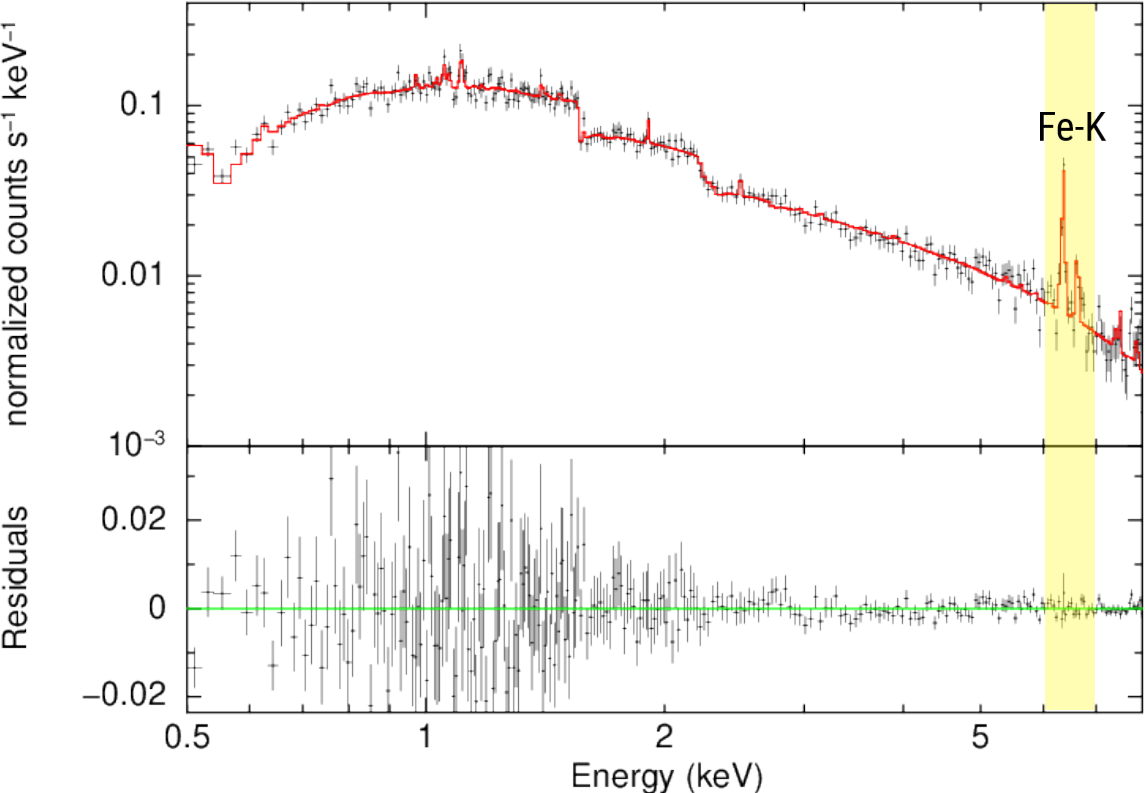}
   \includegraphics[width=0.48\textwidth]{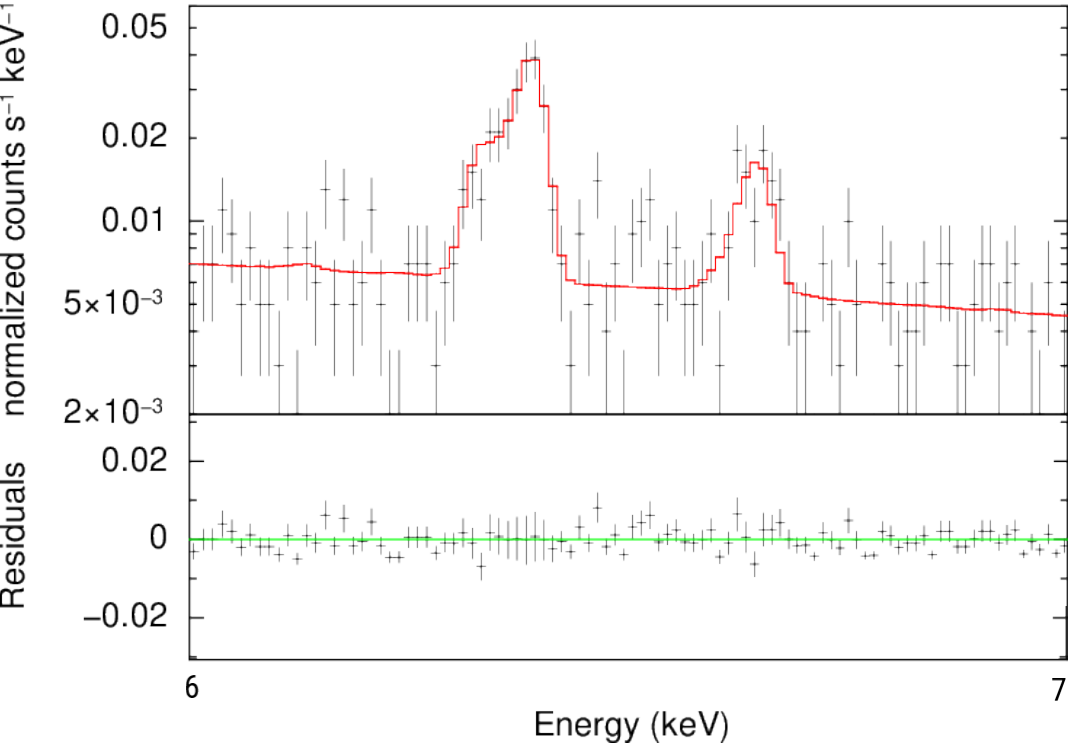} 
   \caption{Typical spectrum of the ICM in equilibrium in our cluster models. Left panel. Fit (upper panel; red line) and residuals (lower panel) to the mock spectrum of the receding ICM (black crosses), in equilibrium in RMS model, for the plane-of-the-sky region R1, taking $\sigmaturb =500\, \mathrm{km/s}$.
   FeXXV and FeXXVI (with rest-frame energies close to 6.7 and $6.9\,\mathrm{keV}$, respectively), referred to as Fe-K, are the prominent emission lines in the upper panel.
   In the lower panel, the green line traces the null residuals.
   We rebinned the data for display purposes (we recall that using the C-Statistics the data without background are not binned). 
   Right panel. Zoom in the range $6-7\,\mathrm{keV}$ of the yellow band in the left panel, where the emission lines FeXXV and FeXXVI are emphasized.
   Here, we rebin the data in a way different from the left panel for display purposes.
   }
    \label{fig.FIT}
    \end{figure*}

At $E> 2 \,\mathrm{keV}$, the strongest and better-modeled lines of the X-ray spectra are due to the transitions of inner shell electrons of the iron in the ICM (see e.g.\  \citealt{Z12,OTA18}, and Fig.\ref{fig.FIT}, where we show a typical spectrum of the ICM, discussed in detail below).
The iron thus represents the reference element for the calculations on the line broadening.
Previous works have shown that, though being formally independent of the line broadening, the best-fitting Doppler shift of X-ray emitting lines is decisively affected by their broadening. 
In particular, on the basis of the results of the fits to mock observations of the rotating ICM, \citet{B13} point out that, at a fixed signal-to-noise ratio, the best-fitting Doppler shift of the centroids of the X-ray emission lines suffers from a higher error when increasing their overall broadening above $\approx 300\,\mathrm{km/s}$.
Such a consideration brings us to take into account the following  contributions to the broadening of the strong iron emitting lines:
\begin{itemize}
    \item The random motion of iron ions produces the thermal broadening ($\sigmaTH$), which is accounted for by the spectroscopic-like temperature (Eq.\ \ref{eq.Tsl}) in BAPEC model. In our mock spectra, $90\,\mathrm{km/s}<\sigmaTH<110\,\mathrm{km/s}$.
    We notice that the adopted value of the spectroscopic-like temperature represents a weighted-average of the observed distribution in the integrated spectra, with typical dispersions around this central value in the range $(0.37-0.53)$ keV for all the models.
    \item The turbulence, which is believed to be ubiquitous in galaxy clusters on the basis of hydrodynamical simulations \citep[e.g.][]{V17} and observations \citep[e.g.][]{S04}, is expected to induce a nonnegligible contribution (in the order of a few hundred km/s) to the broadening of the iron emitting lines, known as turbulent broadening $\sigmaturb$ (e.g. \citealt{Z12}). In the following analysis, we consider a $\sigmaturb$ of both 0 and 500 km/s, the latter one considered as an upper limit on the turbulent velocity dispersion in typical galaxy clusters \citep[see e.g.][]{P15}. 
\end{itemize}

In order to mimic an observation as far as possible realistic, we introduce a typical absorption due to the Milky Way ($N_\mathrm{H}=5\times10^{20} \,\mathrm{cm^{-2}}$; e.g. \citealt{N_H}), using the PHotoelectric ABSorption model\footnote{\href{https://heasarc.gsfc.nasa.gov/xanadu/xspec/manual/XSmodelPhabs.html}{https://heasarc.gsfc.nasa.gov/xanadu/xspec/manual/XSmodelPhabs.html}.} (PHABS).
Assuming also the parameters of Table~\ref{tab.PARA}, an exposure time of 100$\,\mathrm{ks}$, and convolving in the range $0.5$-$8\,\mathrm{keV}$ with instrumental response functions of \emph{Resolve}\footnote{See \href{https://heasarc.gsfc.nasa.gov/docs/xrism/proposals/}{https://heasarc.gsfc.nasa.gov/docs/xrism/proposals/}.} 
in Xspec\footnote{See \href{https://heasarc.gsfc.nasa.gov/xanadu/xspec/}{https://heasarc.gsfc.nasa.gov/xanadu/xspec/}.} (\citealt{Xspec}), we build mock spectra of the rotating ICM of our cluster models (see an example in Fig. \ref{fig.FIT}). 
We do not consider any background in our mock spectra, working in the ideal condition of the analysis of very bright regions. 
To account for the different behavior of response matrices at different energies, for any region under consideration, we present two mock spectra: one for ``approaching''/blue-shifted ICM and another for ``receding''/red-shifted ICM, with typical differences in energy of the line centroids of a few tens of eV (see Fig.~\ref{fig.FIT}). 
Moreover, to assess the impact of the turbulence on the fit to the shape of the emitting lines, for any region under consideration we present a couple of mock spectra: one with $\sigmaturb= 500\, \mathrm{km/s}$ and another without turbulence. 
The emission at 6-$7\,\mathrm{keV}$ (yellow vertical band in the left-hand panel of Fig. \ref{fig.FIT}) provides the most valuable information to measure the l.o.s.\  speed (see also \citealt{OTA18}), because of the relatively high emissivity of iron emitting lines FeXXV and FeXXVI (see  also the right-hand panel of Fig.\ \ref{fig.FIT}). 

Using the C-statistics \citep{Cash79}, as suggested by \citealt{OTA18} \citep[see also][]{Humphrey09,Kaastra17}, and thawing all the parameters except $N_\mathrm{H}$, we then fit the absorbed model BAPEC to the mock spectrum in Fig. \ref{fig.FIT}.
With the purpose of studying the \emph{Resolve} ability to detect the ICM rotation (see Sect. \ref{sec.RECOVER}), in Tab.\ \ref{tab.FIT} we report the expectation values and the statistical errors of the parameters of the fit to the X-ray emission lines: the effective redshift $\zeff$ (that regulates the energy shift of their centroids), the turbulent velocity $\sigmaturb$ (that contributes to their broadening), the metallicity $Z$ (that regulates their intensity), and the spectroscopic temperature $T$ (that is related to a contribution to their broadening).

\begin{table*}
      \caption[]{Results of the spectral analysis of the mock spectra of the ICM. 
      }
         \label{tab.FIT}
     $$ 
         \begin{array}{p{0.13\linewidth}l|llll}
            \hline
            \noalign{\smallskip}
             Model - Region & \sigmaturb[\textrm{km/s}] & \zeff(S_{\rm{eff}}) & \sigmaturb[\mathrm{km/s}] & Z[Z_{\odot}](S_{\rm{Z}}) & T[\mathrm{keV}](S_{\rm{T}})\\
            \noalign{\smallskip}
            \hline
            \noalign{\smallskip}
             RMS - R1-R & 0 & 0.0512\pm 0.0001(0.0) & 1\pm46 & 0.28\pm0.02(1.0) & 6.37\pm0.16(0.0)\\
             RMS - R1-R & 500 & 0.0513\pm 0.0002(0.5) & 478\pm57 & 0.29\pm0.02(0.5) & 6.16\pm0.15(1.3)\\
             RMS - R2-R & 0 & 0.0511\pm 0.0001(0.0) & 4\pm66 & 0.27\pm0.05(0.6) & 5.96\pm0.37(0.1)\\
             RMS - R2-R & 500 & 0.0515\pm 0.0005(0.8) & 522\pm151 & 0.36\pm0.06(1.0) & 6.07\pm0.31(0.2)\\
             RMS - R3-R & 0 & 0.0510\pm 0.0002(0.5) & 3\pm79 & 0.31\pm0.06(0.2) & 5.13\pm0.26(0.7)\\
             RMS - R3-R & 500 & 0.0517\pm 0.0005(1.6) & 826\pm252 & 0.31\pm0.06(0.2) & 5.17\pm0.34(0.6)\\
             \\[0.1mm]
             RMO - R1-R & 0 & 0.0512\pm 0.0001(0.0) & 47\pm43 & 0.28\pm0.02(1.0) & 6.33\pm0.15(0.7)\\
             RMO - R1-R & 500 & 0.0512\pm 0.0002(0.0) & 486\pm49 & 0.30\pm0.02(0.0) & 6.28\pm0.15(0.3)\\
             RMO - R2-R & 0 & 0.0509\pm 0.0001(0.0) & 0\pm79 & 0.33\pm0.05(0.6) & 5.82\pm0.32(1.1)\\
             RMO - R2-R & 500 & 0.0503\pm 0.0005(1.2) & 477\pm123 & 0.26\pm0.04(1.0) & 5.80\pm0.33(1.2)\\
             RMO - R3-R & 0 & 0.0507\pm 0.0002(0.5) & 0\pm105 & 0.31\pm0.06(0.2) & 5.59\pm0.41(1.3)\\
             RMO - R3-R & 500 & 0.0504\pm 0.0004(0.5) & 299\pm161 & 0.38\pm0.07(1.1) & 4.87\pm0.32(0.6)\\
             \\[0.1mm]
             RMP - R1-R & 0 & 0.0515\pm 0.0001(0.0) & 83\pm67 & 0.27\pm0.02(1.5) & 6.53\pm0.18(0.1)\\
             RMP - R1-B & 0 & 0.0485\pm 0.0001(0.0) & 3\pm69 & 0.29\pm0.02(0.5) & 6.83\pm0.20(1.4)\\
             RMP - R1-B & 500 & 0.0484\pm 0.0002(0.5) & 445\pm62 & 0.29\pm0.02(0.5) & 6.49\pm0.17(0.4)\\
             RMP - R2-R & 0 & 0.0514\pm 0.0001(0.0) & 2\pm83 & 0.32\pm0.05(0.4) & 5.36\pm0.29(2.2)\\
             RMP - R2-B & 0 & 0.0486\pm 0.0001(0.0) & 7\pm68 & 0.30\pm0.05(0.0) & 6.21\pm0.38(0.6)\\
             RMP - R2-B & 500 & 0.0483\pm 0.0005(0.6) & 565\pm169 & 0.37\pm0.06(1.2) & 5.72\pm0.34(0.8)\\
             RMP - R3-R & 0 & 0.0512\pm 0.0002(0.5) & 107\pm85 & 0.37\pm0.08(0.9) & 5.86\pm0.49(1.9)\\
             RMP - R3-B & 0 & 0.0489\pm 0.0003(0.0) & 8\pm158 & 0.23\pm0.07(1.0) & 5.20\pm0.44(0.6)\\

            \noalign{\smallskip}
            \hline
         \end{array}
     $$ 
     \tablefoot{Input conditions of the ICM (two columns on the left of the black vertical line): the name of the model of the rotating ICM (RMS, RMO, RMP), the name of the regions where the spectra are integrated (R1, R2 or R3 described in Tab. \ref{tab.REG}; first column), with the positive (identified by -R) or negative (by -B) input $\langle \vlos \rangle$ (first column), and the assumed turbulent velocity dispersion (second column). Output parameters (four columns on the right of the vertical black line) and their statistical errors (at $1\sigma$ of confidence level) of the best fits in our mock spectra of the rotating ICM in equilibrium in model RMS, reporting in brackets the significance of the corresponding "best-fit" results (computed by using Eq. \ref{eq.Q_fit}). }
   \end{table*}

\subsection{Significativity of the recovered observable quantities}
\label{sec.RECOVER} 

In this Section, we discuss how the BAPEC parameters $\zeff$, $\sigmaturb$, $Z$, and $T$ are recovered from the fit of our mock spectra, once convolved with the \emph{Resolve} response matrices in the X-rays. 

We thus introduce the significativity of the "best-fit" quantity $\Qout$ (reported in Tab. \ref{tab.FIT}):
\begin{equation}
\label{eq.Q_fit}
\SsQ=\frac{\left\lvert \Qout-\Qin\right\rvert}{\errQ},
\end{equation} 
where $\Qin$ and $\errQ$ are the input parameter (reported in Tab. \ref{tab.PARA}) and the error of $\Qout$ to $\simeq 68\%$ of confidence (reported in Tab. \ref{tab.FIT}), respectively. 
$\SsQ$ measures at which level of confidence the "best-fit" parameters match the input values: $\SsQ \leq1$ means that the spectral analysis recovers the input parameter $\Qin$ within $\simeq 68\%$ of confidence. 
A lower $\SsQ$ thus corresponds to a better recovery of the observable property $\Qin$ via the spectral best-fitting. 
Using $Q=\{T,\zeff,Z\}$ in Eq. (\ref{eq.Q_fit}), we estimate their significance, reported in Tab. \ref{tab.FIT}, where we refer to the significance of $\zeff$ as $S_\mathrm{eff}$. 
The input parameters of the spectroscopic temperature, metallicity, and effective redshift in most spectral analyses are recovered within $1 \sigma$ of confidence level. 
To illustrate the results of these mock observations, we focus on the best and worst recoveries of the rotation speed of the ICM. 
First, we compare the effective redshift measured in R1 region of RMS cluster model with receding, nonturbulent ICM (see the third column and the first row of Tab.~\ref{tab.FIT}) to the corresponding input $\zeff$ (see the fifth column and the first row of Tab.~\ref{tab.PARA}): 
the output $\zeff$ perfectly matches the input $\zeff$ (i.e. $S_\mathrm{eff}=0$, using Eq.~\ref{eq.Q_fit}).
Second, we compare the effective redshift measured in the region R3 of the RMS cluster model with receding, turbulent ICM (see the third column and the sixth row of Tab.~\ref{tab.FIT}) to the corresponding input $\zeff$ (see the fifth column and the third row of Tab.~\ref{tab.PARA}): 
the output $\zeff$ matches the input $\zeff$ at  $1.6\sigma$.
Though each measurement depends on the signal-to-noise ratio, this exercise shows the ability of \emph{Resolve} to measure the rotation speed of the ICM at high significance, assuming that the cluster cosmological redshift and Milky Way absorption are known. We note that the statistical errors associated to the "best-fit" spectroscopic temperature, effective redshift, and metallicity (Tab.\ \ref{tab.FIT}) depend on the signal-to-noise ratio: these errors decrease by raising the signal-to-noise ratio, i.e.\ by increasing the exposure time (here assumed to be 100 ks) and enlarging the plane-of-the-sky exposure region (see Tab.~\ref{tab.REG}).
For instance, comparing the RMS-R3-R spectral analyses with $\sigmaturb=0$ and $\sigmaturb= 500 \,\mathrm{km/s}$ (see Tab.~\ref{tab.FIT}), we note that, keeping fixed the signal-to-noise ratio, the increase in $\sigmaturb$ induces a higher error of "best-fit" effective redshift and turbulent velocity.
Most importantly, in this case the input $\zeff$ is recovered within $1\sigma$ if input $\sigmaturb=0$ and out of $1\sigma$ if input $\sigmaturb= 500 \,\mathrm{km/s}$.
From the entire set of our results, the significativity of effective redshift appears to be sensitive to the input turbulent velocity dispersion: the spectral best-fitting recovers, on average, the input $\zeff$ with higher $S_{\rm{eff}}$ (i.e. within a higher confidence level), when we increase the input $\sigmaturb$.
This outcome is in line with the picture that emerged from the X-ray mock observations of galaxy clusters from hydrodynamical simulations, where the increase in the complexity of the velocity field (here, obtained with increasing turbulent velocity dispersion, at fixed rotation speed) reduces our ability to recover the kinematic properties of the ICM \citep[e.g.][]{R18}.

We have also studied the covariance among the BAPEC best-fit parameters.
We obtain that the off-diagonal correlation coefficients are significantly lower than 0.2, implying no relevant cross-correlation, between $\zeff$, $\sigmaturb$, $Z$, and $T$.
A partial exception is the $\approx 0.2$ correlation coefficient between $Z$ and $\sigmaturb$ for all the models: this weak correlation is due to the way $Z$ is measured ($Z$ is estimated by measuring the equivalent width of emitting lines).
In conclusion, we find that the cross-correlations have a negligible impact on our measurements of the ICM rotation speed.

Using the configurations for \emph{Resolve}, we conclude that, even in the presence of the turbulence of $500\, \mathrm{km/s}$, the l.o.s.\ component of the rotation velocity is recovered through the fitting procedure within $1\sigma$ of confidence level in most analyses of the mock spectroscopic data.
The analysis of our cool-core cluster models shows that current observational constraints, such as the rotation speed of the ICM based on the upper limits on the broadening of the X-ray emitting lines, the measurements of the thermodynamic profiles, the flattening of the surface brightness distribution and of the hydrostatic mass bias, leave room for rotation of the ICM up to $600\,\mathrm{km/s}$ in typical clusters. Further tests of our cluster models with rotating ICM will be provided by future measurements of the l.o.s.\ velocity with \emph{XRISM/Resolve} that will put stringent and direct constraints on the intrinsic kinematics of the ICM in galaxy clusters.

\subsection{Assessing the hydrostatic mass bias with X-ray spectroscopy}
\label{sec.REDUCE}

In our cluster models, the ICM is in equilibrium and departs from the hydrostatic condition owing only to rotation.
Here, we point out the perspectives and limitations on the use of X-ray spectroscopy for the mapping of nonnegligible rotation support of the ICM.

As discussed above, the l.o.s.\ velocity $\vlos(x,z)$ (see Eq. \ref{eq.vlos}) can be recovered from the measurements of the properties of the X-ray emitting lines \citep[see e.g.][]{biffi13,R18}.
Thus, a proxy for the rotational contribution to the hydrostatic mass bias, defined in Eq.\ (\ref{eq.BHE}), is $\Mrot/\Mtrue$, where 
\begin{equation}
    \label{eq.Mrot}
    \Mrot(<r)=\frac{\vlos^2(r,0)r}{G}
\end{equation}
is the mass associated with the gas rotation support, and $\Mtrue$ the same halo mass as in Eq. (\ref{eq.BHE}). 

\begin{figure}
   \centering
   \includegraphics[width=0.49\textwidth]{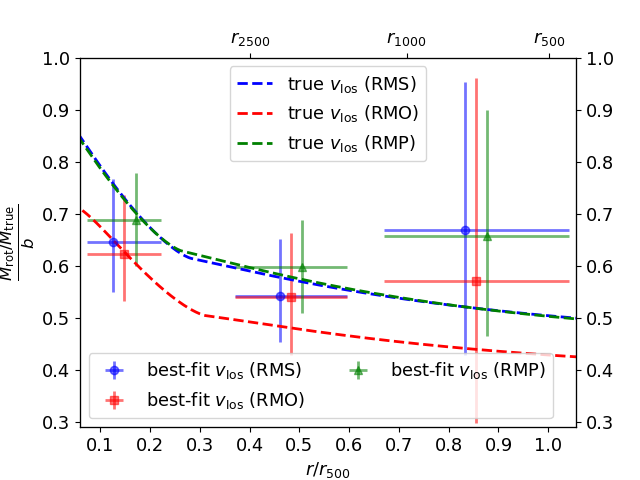}
   \caption{
   Fraction of the hydrostatic mass bias due to rotation ($b$) of RMS (blue), RMO (red) and RMP (green) models that we recover from the true (dashed lines) and best-fit (points) l.o.s.\ velocities using Eq.\ (\ref{eq.Mrot}). 
   The horizontal error bars indicate the extent of the region of the mock observations, while the vertical error bars the $1\sigma$ errors on $\Mrot$ estimated from X-ray spectroscopy.
   Radius is normalized to $\rcinque$ as in Fig. \ref{fig.b_all}, with the RMS radii $r_{2500}$, $r_{1000}$ and $r_{500}$ indicated in the top axis.
   }
   \label{fig.b_fit}
    \end{figure}

Using in Eq.~(\ref{eq.Mrot}) the true l.o.s.\ rotation speed $\vlos$, given by Eq. (\ref{eq.vlos}), we compute the $\Mrot$ profiles of our cluster models (see curves in Fig. \ref{fig.b_fit}). 
Then, to find the l.o.s.\ velocity $\langle \vlos\rangle$ as measured from the best-fits to our mock spectra, we use Eq. (\ref{eq.zeff}), where $\zeff$ is now the best-fit value to the mock spectrum of the receding ICM without turbulence (reported in Tab. \ref{tab.FIT}). 
Substituting $\langle \vlos\rangle$ instead of $\vlos(r,0)$ in Eq. (\ref{eq.Mrot}), where we consider the radius $r$ equal to the value of the plane-of-the-sky $x$-coordinate (reported in Tab. \ref{tab.REG} for the region under consideration), we estimate the mass associated with gas rotation support at the centers of the regions chosen for our mock observations.
Following this method, from the normal distribution with mean and standard deviation equal to the best-fit effective redshift and its error (both reported in Tab. \ref{tab.FIT}), respectively, we infer the errors (as 16th and 84th percentiles) on $\Mrot$ as estimated from X-ray spectroscopy for our cluster models.

Fig. \ref{fig.b_fit} shows that $\Mrot$ estimated from the best-fit $\zeff$ recovers within $1\sigma$ statistical errors (vertical error bars) the mass associated with the rotation support estimated from the true l.o.s.\ velocity (Eq. \ref{eq.vlos}). 
This is consistent with the fact that the best-fit effective redshift from the spectral analysis recovers within $1\sigma$ of confidence level the input effective redshift (Sect. \ref{sec.RECOVER} ).
However, as shown by the curves in Fig.~\ref{fig.b_fit}, $\Mrot/\Mtrue$ based on Eq. (\ref{eq.Mrot}), where we take the true l.o.s.\ speed, is lower than the hydrostatic mass bias $b$, measured from the theoretical angle-averaged pressure profile of the ICM (see Sect. \ref{sec:bias}). 
The reason for this discrepancy \cite[pointed out also by][]{OTA18} is that the mean l.o.s.\ velocity at a projected distance $d$ from the symmetry axis is lower than the rotation speed of the ICM at an intrinsic distance $d$ from the symmetry axis.

Focusing on the hydrostatic mass biases of our cluster models as measured from X-ray spectroscopy (points with error bars in Fig.\ \ref{fig.b_fit}), we conclude that the estimates of the rotation support over the range $(0.1-1) \rcinque$ obtained through the "best-fit" l.o.s.\ rotation speed resolved by \emph{Resolve} are able to account for $55-70\%$ of the hydrostatic mass bias of our models.
It follows that a \emph{Resolve}-like correction for the rotation support of the ICM is expected to leave a residual hydrostatic mass bias due to rotation smaller than $3\%$ at $\rcinque$ for systems similar to our model clusters.
Error bars in Fig.~\ref{fig.b_fit} are larger in the outermost bin for all the models: this is a consequence of the increase of the statistical uncertainties of the spectral parameters due to the lower signal present in those regions of our cluster models.

Moreover, the poor angular resolution of \emph{Resolve} (with a Point-Spread-Function with a half power diameter of $\approx 1.7'$) prevents us from sampling the hydrostatic mass bias profile in a larger independent number of radial bins.  
This will be possible in the future with the Advanced Telescope for High Energy Astrophysics\footnote{The ESA satellite \emph{ATHENA}, is scheduled to be launched not before 2036 (see \href{https://www.the-athena-x-ray-observatory.eu/en}{https://www.the-athena-x-ray-observatory.eu/en}).} (\emph{ATHENA}; \citealt{Nandra13}), thanks to its expected arcsec resolution combined with the performance of the onboard X-ray microcalorimeter \emph{X-IFU} \citep[see e.g.][]{R18}.

\subsection{Discussion on the hydrostatic bias}
\label{sec.DISC}

In Section~\ref{sec:bias}, we discussed how the measurements of the hydrostatic mass bias can be used to limit the rotation speed of the ICM.
In Table~\ref{tab.BIAS}, we quote the hydrostatic mass bias due to rotation in our cluster models at some characteristic overdensities available to observations. 
A general trend is that the observed hydrostatic mass bias decreases with increasing overdensity \citep[see e.g.][]{ZHANG10,MAHDAVI2013,SERENO15,lovisari20}.
A similar trend is also recovered in hydrodynamical simulations \citep[see e.g.][]{N07,lau09,meneghetti10,rasia12,GIANFAGNA21}.
This behavior results in tension with the hydrostatic mass bias profiles recovered from our models, which increase with increasing $\Delta$ (see Fig.~\ref{fig.b_all}). 
Cosmological hydrodynamical simulations show that the support from turbulence in galaxy clusters increases with radius \citep[see e.g.][]{F09,lau09,Towler23}, overcoming the rotational contribution well within $r_{500}$.
Thus, the observed trend of the hydrostatic mass bias is expected to follow the increase in the turbulent support of the ICM moving outward, 
with a non-negligible contribution from the rotation only in the inner regions. Indeed, the few data available at $r_{2500}$ (see Fig.~\ref{fig.b_all}), where hydrodynamical simulations suggest comparable support from rotation and turbulence, suggest a hydrostatic bias marginally consistent (within $2\sigma$) with the predictions of  our models.

In the near future, new instruments and space telescopes will permit more accurate determinations of the hydrostatic mass bias at different overdensities in a larger sample of galaxy clusters.
In particular, the aforementioned \emph{XRISM} and \emph{eRosita}\footnote{See \href{https://www.mpe.mpg.de/eROSITA}{https://www.mpe.mpg.de/eROSITA}.} (onboard the Spectrum-Roentgen-Gamma mission and, only in the future, the observatory \emph{Athena}), together with currently available X-ray observatories (\emph{XMM-Newton} and \emph{Chandra}\footnote{See \href{https://chandra.harvard.edu/}{https://chandra.harvard.edu/}.}), will continue to provide the measurements of the hydrostatic mass through X-ray observations. 
The ESA optical/infrared space telescope \emph{Euclid}\footnote{See \href{{https://sci.esa.int/web/euclid}}{https://sci.esa.int/web/euclid}.} and other ground-based campaigns will complement with weak lensing mass estimates the information on the mass budget in larger samples of galaxy clusters, allowing us to refine our comprehension of the statistical properties of the hydrostatic mass bias.

\begin{table}
\caption[]{Characteristic values of the hydrostatic mass bias of our cluster models.
}
\label{tab.BIAS}
\centering          
\begin{tabular}{cccc}
\hline
            \noalign{\smallskip}
            \multirow{2}{1em}{$\Delta$} & \multicolumn{3}{c}{$b$}\\
            & RMS & RMO & RMP\\
\hline
            \noalign{\smallskip}
            2500 & 0.09 & 0.07 & 0.13\\
            1000 & 0.06 & 0.04 & 0.09\\
            500 & 0.05 & 0.03 & 0.07\\
            \noalign{\smallskip}
            \hline                
\end{tabular}
\tablefoot{We quote the hydrostatic mass biases ($b$) of our clusters models (RMS, RMO and RMP) at $\rDelta$ with $\Delta=\{2500, 1000, 500\}$.}
\end{table}

\section{Conclusions}
\label{sec.CONCL}
In this work, we have presented three representative, realistic models of massive ($\Mdue \approx 10^{15}\, \mathrm{M_\odot}$) cool-core galaxy clusters with rotating ICM in equilibrium in dark matter halos consistent with observational findings and theoretical predictions on the halo shape and mass-concentration relation (Sect.\ \ref{sec.2}). 
While one of the models has a spherical NFW halo, the other two have, respectively, physically consistent oblate and prolate NFW halos, built analytically using the method of \cite{C05}.
Our cool-core cluster models, which have barotropic ICM rotation with velocity peaks as high as $600\,\mathrm{km/s}$ (see Fig. \ref{fig.RC}), have ICM temperature and density profiles consistent with the corresponding universal profiles of real clusters. 
Cosmological hydrodynamical simulations can also be used to calibrate these analytic models (for instance, on the location of $R_{\rm break}$, the parameter that defines the size of the cool core) once any overcooling problem \citep[see e.g.][]{K12} is properly solved, and realistic cooling cores are produced in systems that did not experience a major merger in the central region \citep[e.g.][]{R15}.
The shape of surface-brightness contours, the discrepancy between hydrostatic and true masses, and 
the broadening of X-ray emission lines of the models are also consistent with currently available observations.

We obtained a set of mock X-ray spectra of the rotating ICM from the aforementioned three cluster models, using the configuration for the microcalorimeter \emph{Resolve} onboard \emph{XRISM}, for different turbulence conditions.
In this way, we estimated how well the rotation speed and the hydrostatic mass bias due to rotation are recovered based on the results of \emph{Resolve}-like spectral analysis (Sect.\ \ref{sec.DETECT}).

The main conclusions of this work are the followings:
\begin{itemize}
    \item The existence of realistic cluster models with the peaks of the rotation speed of the ICM in the range 400-$600\,\mathrm{km/s}$ leaves open the possibility that the rotation support of the ICM is nonnegligible in real cool-core galaxy clusters. 
    \item Even with turbulent velocity dispersion as high as $500\,\mathrm{km/s}$, a \emph{Resolve}-like X-ray spectral analysis recovers the input l.o.s.\ rotation speed at high significance. 
    \item 
    Measuring the \emph{}{line-of-sight} velocity from X-ray spectroscopy with \emph{XRISM} accounts for $55-70\%$ of the hydrostatic mass bias due to rotation.
    In this way, \emph{XRISM} will allow us to pin down any mass bias of origin different from rotation (for instance, due to turbulence; see e.g. \citealt{E22}).
\end{itemize}

On one side, improving spatial and spectral resolution in X-rays will open a new window in which the combination of the intrinsic thermodynamic profiles with the rotation and turbulent velocity dispersion profiles can be used to validate models of the ICM, providing robust estimates of the cluster mass. 
On the other side, Sect.~\ref{sec.REDUCE} shows the need for a functional form that properly maps the intrinsic rotation speed through the line-of-sight rotation speed as resolved in massive clusters. Most of the limitations of this mapping come from the possible degeneracy present in the interpretation of the observational data. 
Possible contaminants that can limit our interpretation of the physical state of the ICM are, for example, unresolved gas clumps, multiphase gas, metallicity inhomogeneities, and complex velocity fields not properly mapped both in the plane of the sky and along the line of sight (see also Sect. \ref{sec.RECOVER}). We postpone further study on this topic to future work.

X-ray observations will enable us to guess both the rotation axis and the maximal rotation speed \citep[see e.g.][]{OTA18,L19} in some favorable conditions (broadly speaking, bright enough source and X-ray detector with sufficient spatial and spectral resolution).
Once these X-ray observations are available, the kinetic Sunyaev-Zeldovich \citep[see e.g. some observational constraints in][for a review]{sayers13,sayers19,mroczkowski19} can be resolved (thanks also to the forthcoming ground-based Simons Observatory; \citealt{Ade19}) and compared to the X-ray constraints to provide a consistent picture of the ICM peculiar velocity along the line of sight.

The presented results strongly encourage future spectroscopic observations of relaxed galaxy clusters with \emph{XRISM/Resolve} (in the forthcoming decade) and/or \emph{ATHENA/X-IFU} (in the far future; see also \citealt{R18}) to quantify the level of the ICM rotation speed, and to reduce as far as possible the hydrostatic mass bias in real clusters, with important implications for the use of galaxy clusters as accurate cosmological proxies \citep[see e.g.][]{P19}. 

As pointed out by \citet{Nipoti2014} and \citet{N15}, if the ICM is weakly magnetized (as found by the observational works reviewed by \citealt{BRU13}) and significantly rotating, the magnetorotational instability could also have relevant effects.
Thus, the possibility that the ICM has nonnegligible rotation support with a speed as high as $600\,\mathrm{km/s}$ in real clusters acquires a great interest for the implications not only on the mass estimates, which are needed to use galaxy clusters as cosmological probes, but also for our understanding of the energy balance and evolution of the cool cores, because the magnetorotational instability could play a role in regulating their energetic budget.

\begin{acknowledgements}
We thank the referee Edoardo Altamura for useful suggestions.
S.E.\ acknowledges the financial contribution from the contracts ASI-INAF Athena 2019-27-HH.0,
``Attivit\`a di Studio per la comunit\`a scientifica di Astrofisica delle Alte Energie e Fisica Astroparticellare''
(Accordo Attuativo ASI-INAF n. 2017-14-H.0), and from the European Union’s Horizon 2020 Programme under the AHEAD2020 project (grant agreement n.\ 871158).
\end{acknowledgements}

\bibliography{refs}{}
\bibliographystyle{aa}

\begin{appendix}

    \section{An extreme cluster model with rotating ICM}
    \label{A.THERMO}

\begin{table*}
      \caption[]{Parameters of "rotating model extreme" (RME) model.}
         \label{tab.EXT}
     $$ 
         \begin{array}{p{0.2\linewidth}llllllll}
            \hline
            \noalign{\smallskip}
            Model & {\rm Halo} & \Rbreak[\mathrm{kpc}] & \neestar[\mathrm{cm^{-3}}] & \Tstar[\mathrm{keV}] & \uzero[\mathrm{km/s}] & \Rzero[\mathrm{kpc}] & \gammain & \gammaout\\
            \noalign{\smallskip}
            \hline
            \noalign{\smallskip}
            RME & {\rm DMO} & 420 & 2.5\times 10^{-3} & 7.8 & 3000 & 100 & 0.60 & 1.55\\
            \noalign{\smallskip}
            \hline
         \end{array}
     $$ 
     \tablefoot{Same as Tab. \ref{tab.ROT}, but for model RME.}
   \end{table*}

\begin{figure}
   \centering
   \caption{Thermodynamic (upper and central panels) and rotation speed profiles (lower panel) of the ICM in "rotating model extreme" (RME) model. Upper and central panels. Same as the left and right panels of Fig. \ref{fig.RMS}, respectively, but for model RME. Lower panel. Rotation speed profile of the ICM in model RME. The red arrow indicates approximately the extent of the cool core, defined as in Fig. \ref{fig.RC}.}
   \includegraphics[width=0.49\textwidth]{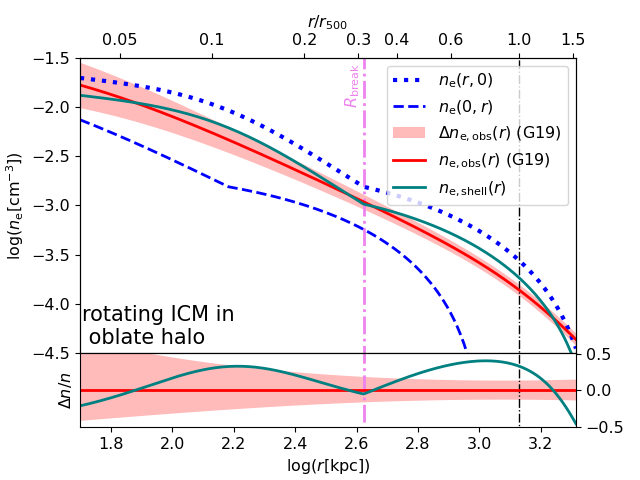}
   \includegraphics[width=0.49\textwidth]{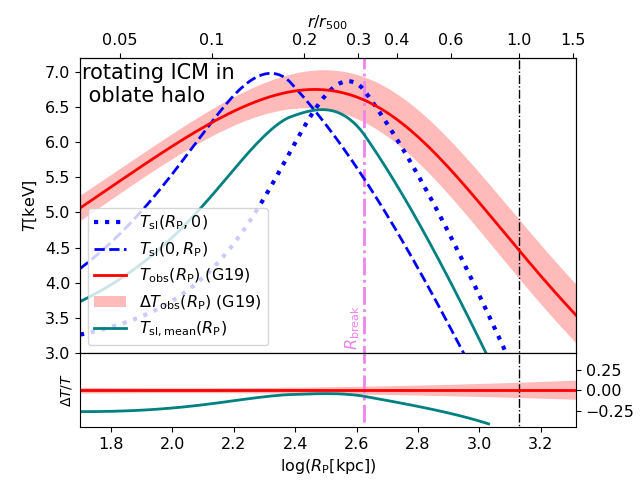}
   \includegraphics[width=0.49\textwidth]{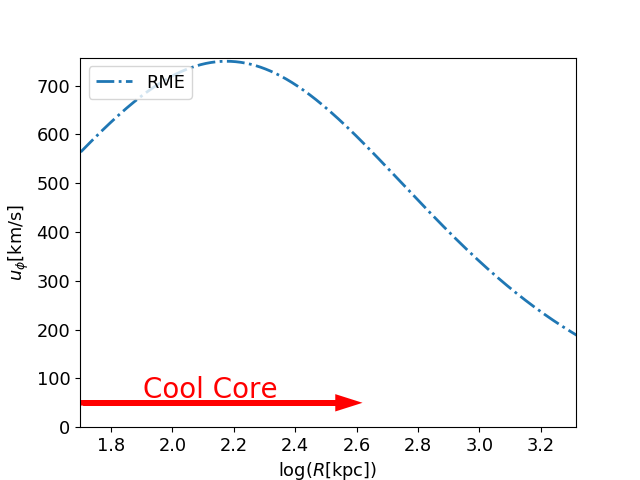}
    \label{fig.EXT}
    \end{figure}

In this appendix, with the purpose of illustrating the effect of strong ICM rotation on observable properties of galaxy clusters, we present a cluster model (of $\Mdue \approx 10^{15}\,\mathrm{M_\odot}$) with rotating ICM, which, different from the three models presented in Sect.~\ref{sec.ROT}, is not realistic,
because, though having realistic gas density distribution, has a temperature distribution substantially different from that of real clusters. 
This extreme cluster model, which we refer to as "rotating model extreme" (RME), has gravitational potential generated by the oblate halo model DMO (see Sect.\ \ref{sec.2.2}) and gas rotation law given by Eq.~(\ref{eq.RE4}) with values of the parameters $\uzero$ and $\Rzero$ (see Tab.\ \ref{tab.EXT}) such that rotation speed peak is $\simeq 750\,\mathrm{km/s}$ at a radius $\simeq160\,\mathrm{kpc}$ (see bottom panel of Fig.\ \ref{fig.EXT}).
The values of the other gas parameters ($\Rbreak$, $\gammain$, $\gammaout$, $\neestar$ and $\Tstar$; see Tab.\ \ref{tab.EXT}) are chosen so that the angle-averaged gas density profile of model RME is consistent with the universal gas density profile of observed cool-core clusters (top panel of Fig.\ \ref{fig.EXT}).
However, this choice of the values of the parameters implies that, due to the strong rotation support, the temperature profile of model RME is grossly inconsistent with the universal temperature profile derived for observed cool-core clusters (middle panel of Fig.\ \ref{fig.EXT}).

Given the rotation speed curve and the gravitational potential assumed for model RME, we were not able to find a combination of values of the plasma parameters such that both the density and the spectroscopic-like temperature profiles are consistent with those observed for massive cool-core clusters. 
Though this does not allow us to place an upper limit on the peak of the rotation speed of the ICM, it is a strong indication that rotation speeds higher than $\approx 600$ km\,s$^{-1}$ are problematic not only for the spectroscopic constraints on the broadening of the X-ray emission line, but also for constraints imposed by the shape of the universal thermodynamic profiles. 
Comparing further the polytropic indices $\gammain$ and $\gammaout$ of our ICM distributions, to those observed, we also note that model RME, which requires lower $\gammain$ and higher $\gammaout$ than our realistic models (see Tab.s \ref{tab.ROT} and \ref{tab.EXT}), is in tension with the results of \cite{GPOLY} on the polytropic indices of observed clusters (see Sect.\ \ref{sec.THERMO}).

\end{appendix}

\end{document}